\documentclass[11pt,a4paper]{article}
\usepackage[latin1]{inputenc}
\usepackage{mathtools}
\usepackage{amsfonts}
\usepackage{amssymb,bm}
\usepackage{graphicx}
\usepackage[scale=0.775]{geometry}
\usepackage[dvipsnames]{xcolor}
\usepackage{multirow}
\usepackage{array}
\usepackage{booktabs}
\usepackage{hyperref}
\usepackage{pdflscape}
\usepackage{subcaption}

\newcommand{\vv}{{\bm{v}}}
\newcommand{\ff}{{\bm{f}}}
\newcommand{\ppsi}{{\bm{\psi}}}
\renewcommand{\AA}{{\bm{A}}}

\graphicspath{{./Figures/TESTS/}}

\title{Implicit relaxed all Mach number schemes for gases and compressible materials
}
\author{
	Andrea Thomann\footnote{Institut f\"ur Mathematik, Johannes-Gutenberg-Universit\"at Mainz, Staudingerweg 9, 55099 Mainz, Germany} \footnote{Correspondence to: Andrea Thomann. Email: athomann@uni-mainz.de},
	Angelo Iollo\footnote{Universit\'e Bordeaux, IMB UMR 5251, F-33400 Talence, France; and Equipe-projet Memphis, Inria Bordeaux-Sud Ouest, F-33400 Talence, France (angelo.iollo@inria.fr)},
	Gabriella Puppo\footnote{Dipartimento di Matematica, La Sapienza Universit\`a di Roma, Piazzale Aldo Moro 5, 00185 Roma, Italy (gabriella.puppo@uniroma1.it)}
}
\date{\today}
\begin{document}
	\maketitle
	
	% equation numbers
	\numberwithin{equation}{section}
	\section*{Abstract}
	We present an implicit relaxation scheme for the simulation of compressible flows in all Mach number regimes based on a Jin-Xin relaxation approach.
    The main features of the proposed scheme lie in its simplicity and effectiveness. 
    Thanks to the linearity of the flux in the relaxation system, the time-semi discrete scheme can be reformulated in linear decoupled elliptic equations resulting in the same number of unknowns as in the original system. 
    To obtain the correct numerical diffusion in all Mach number regimes, a convex combination of upwind and centered fluxes is applied. 
    The numerical scheme is validated by applying it on a model for non-linear elasticity. 
    Simulations of gas and fluid flows, as well as deformations of compressible solids are carried out to assess the performance of the numerical scheme in accurately approximating material waves in different Mach regimes. 

	\textbf{Keywords:} All-speed scheme, relaxation method, low Mach number limit, Eulerian elasticity, linearly implicit higher order schemes
	\section{Introduction}
    Various physical phenomena are affected by drastic changes in the speed of sound, or more generally, in the speed of certain waves. 
    The occurrence of these changes may be caused, for example, by the geometry of the problem. 
    Other examples are the propagation of waves in heterogeneous compressible solids.
    These waves can propagate with different velocities due to the local stiffness of the material.
    
    This work focuses on numerical schemes for all Mach number flows in both fluid dynamics and continuum mechanics. 
    Phenomena of interest involve fluid flows and elastic materials whose deformations are investigated within a monolithic Eulerian framework. 
    With this approach any material (gas, liquid or solid) can be described with the same system of conservation equations and a suitable general formulation of the constitutive law \cite{de2016cartesian,godunov2013elements,gorsse2014simple,plohr1988conservative,plohr1992conservative}. 
    This is in contrast to heterogeneous models which switch between different systems and type of equations to describe the evolution of fluids/gases and solids.
    The classical Euler system used in fluid dynamics can be considered as a special case within this general framework. 
    
    The challenges in constructing an all-speed scheme that can be applied for compressible as well as low Mach schemes are twofold. 
    Firstly the scheme has to ensure the correct numerical viscosity for all Mach number regimes. 
    Explicit upwind schemes in particular are not suited for applications in weakly compressible regimes, since they suffer from excessive diffusion when the Mach number is small \cite{GuillardViozat1999,Klein1995} leading to spurious solutions.
    Here, the excess of viscosity is observed with respect to fluctuations in the stress tensor which lead to both acoustic compression and elastic deformations. 
    Having the correct numerical diffusion is an integral part of obtaining so called asymptotic preserving schemes. 
    It is well known that compressible flow equations formally converge to incompressible equations when the Mach number tends to zero, see e.g. \cite{Dellacherie2010,KlainermanMajda1981,Schochet2005} and references therein. 
    To obtain physically admissible solutions especially in the weakly compressible flow regime, the numerical scheme has to preserve these asymptotics.
    Secondly, explicit schemes have to obey a severe time step restriction to guarantee the robustness of the numerical scheme.
    Being forced to use small time steps reflects in long CPU times, in particular if long time intervals are considered. 
    As a side effect all waves are resolved even though usually only the slow waves are of interest and inaccuracies or a lack of resolution on fast longitudinal or acoustic waves are acceptable.
    Therefore, the ability to use large time steps is desirable. 
    This can be achieved with implicit-explicit (IMEX) schemes or fully implicit schemes.
    A non-exhaustive list of more recent IMEX schemes is for example given by \cite{AbgTor2020,AvgBerIolRus2019,BoscarinoRussoScandurra2018,BouchutFranckNavoret2020,ZeifangEtAl2020} and of implicit schemes \cite{AbbIolPup2017,AbbIolPup2019,BerthonKlingenbergZenk2018,viozat1997} as well as references therein.
    The problematic of stability in connection with flux splittings to construct IMEX schemes is addressed for example in \cite{SchNoe2015,ZakNoe2018} which can be avoided using fully implicit schemes. 
    However, IMEX and implicit schemes have to be constructed carefully to avoid the necessity of utilizing non-linear implicit solvers. 
    
    The schemes we propose here are motivated by the work given in \cite{ AbbIolPup2017} where a Jin-Xin relaxation method \cite{JinXin1995} is used to build a linear diffusive approximation to the original equations. 
    In \cite{ AbbIolPup2017}, a linearly fully implicit all-speed integration method for compressible materials is proposed, at the cost of introducing additional auxiliary variables that have to be updated within the scheme. 
    This scheme was found to be accurate in computing steady state solutions as well as in approximating material waves in various Mach regimes and different materials. 
    It showed consistent improvement in approximating material waves at slow velocities with respect to explicit schemes. 
    However it proved to be quite computationally costly due to the complexity of the coupled linear system that needs to be solved implicitly.
    Having to store an increased number of variables is in particular problematic for large scale or multi-dimensional simulations.  
    
    Here, we overcome the need to solve for auxiliary variables while preserving the properties of the linearly implicit scheme proposed in \cite{ AbbIolPup2017}.
    To achieve this, we split the stiff relaxation source terms from the fluxes and then reformulate the homogeneous part in an elliptic form.
    This technique is also used in the context of constructing all-speed schemes for the Euler equations \cite{CordierDegondKumbaro2012,NoelleEtAl2014,ThoZenkPupKB2020}.
    This procedure yields a decoupled symmetric linear implicit system that is easy to implement and can be solved efficiently using standard linear iterative or direct solvers. 
    To reduce the numerical diffusion of the first order scheme, we extend the time semi-discrete scheme to second order. 
    In \cite{CoulFraHelRatSon2019} a Crank-Nicholson scheme \cite{CraNich1947} was used to obtain an implicit second order relaxation scheme for compressible flow only. 
    Here, we use a second order stiffly accurate diagonally implicit Runge Kutta integrator \cite{HaiWan1991} which has an additional computational stage but proved to be stable also for low Mach flows.  
    In space a convex combination of upwinding and centred differences depending on the local Mach number, as in \cite{AbbIolPup2017}, is used which leads to the correct numerical viscosity of the scheme which is validated by several numerical tests. 
    Especially for low Mach flows the centred discretization of the pressure gradient is recovered whose importance for the correct limit behaviour was discussed e.g. in \cite{Dellacherie2010,GuillardViozat1999}.
    
    The paper is organized as follows.
    In Section \ref{sec:NumScheme} we shortly revisit the Jin-Xin relaxation technique described in \cite{JinXin1995} for general systems of hyperbolic conservation laws. For simplicity and illustrating purposes, all derivations are carried out in a one-dimensional framework. 
    Then the numerical scheme based on this relaxation model is developed.
    The derivation of the scheme is kept quite generic and can be applied to general hyperbolic conservation laws.
    First a first order semi-discrete scheme using backward Euler is constructed which is then equipped with a suitable space discretization. 
    To reduce the diffusiveness of the backward Euler scheme, a second order time integrator is employed in the subsequent section. 
    The first and second order schemes are numerically validated in Section \ref{sec:Numerics} applied to the Eulerian model of nonlinear elasticity that is briefly described in Section \ref{sec:NonlinElast}.
    It is tested in different Mach number regimes applied to different materials.
    The results are compared to a standard local Lax-Friedrichs scheme. 
    Final conclusions are drawn in Section \ref{sec:conclusions}. 
    
    \section{Construction of the implicit Jin-Xin relaxation scheme}
    \label{sec:NumScheme}
    We consider a general system of hyperbolic conservation laws, for simplicity in one space direction, given by 
    \begin{equation}
        \label{Sys:Hyperbolic_1D}
        \partial_t \ppsi + \partial_x \ff(\ppsi) = 0, 
    \end{equation}
    where $\ppsi \in \mathbb{R}^k$ denotes the state vector consisting of $k$ variables and $ \ff(\ppsi) \in \mathbb{R}^k$ denotes a (non-linear) flux function. 
    
    Especially concerning the simulation of low Mach flows, stiff terms arise in the flux function which makes the resolution of those flow regimes quite challenging. 
    In combination with the appearance of large characteristic speeds this can cause problems in the construction of numerical schemes applicable on those flows.  
    Here, the aim is to obtain efficient and robust numerical procedures independently of the considered system of hyperbolic conservation laws. 
    To achieve this, we use the Jin-Xin relaxation method \cite{JinXin1995} to build a linear diffusive approximation to the original equations \eqref{Sys:Hyperbolic_1D} called the \textit{relaxation model}. 
    By doing this, we introduce dissipation errors while, due to the linearity of the relaxation model, the numerical solver does not require sophisticated or complicated non-linear solvers.
    Especially regarding stiff gradients, this would increase the complexity of the numerical scheme. 
    
    Motivated by this, we shortly describe the relaxation method in the next section following the work of \cite{JinXin1995}.
    The numerical scheme is constructed subsequently. 
    
    \subsection{The relaxation method}
    \label{sec:JXrelax}
    
    Following \cite{JinXin1995}, we introduce a vector of relaxation variables $ \vv\in \mathbb{R}^k$ and consider the following relaxation system given by 
    \begin{subequations}
        \label{Sys:Relaxation_1D}
        \begin{align}
            \partial_t \ppsi + \partial_x \vv &= 0, \label{eq:Relax_psi}\\
            \partial_t \vv + \AA^2 \partial_x \ppsi &= - \frac{1}{\eta} \left(\vv - \ff(\ppsi)\right), \label{eq:Relax_v}
        \end{align}
    \end{subequations}
    where $\eta > 0$ denotes the relaxation rate and $\AA^2$ is a diagonal matrix with positive entries given by 
    \begin{equation}
        \AA^2 = \text{diag}(a_1^2, \dots, a_k^2). 
    \end{equation}
    The fluxes on the relaxation model are linear whereas the relaxation source term on the right hand side of \eqref{eq:Relax_psi} is non-linear and stiff for small $\eta$.
    Rewriting equation \eqref{eq:Relax_v}, using a Chapman-Enskog expansion of the relaxation variables for small $\eta$, we find 
    \begin{equation}
        \label{eq:Relax_v_exp}
        \vv = \ff(\ppsi) - \eta (\bm f^\prime(\ppsi)^2 - \AA^2)) \partial_x \ppsi,
    \end{equation}
    where $\bm f^\prime(\ppsi) = \nabla_\ppsi \ff(\ppsi)$ denotes the Jacobian of the flux function $\bm f$. 
    We have neglected the $\mathcal{O}(\eta^2)$ terms in the expansion \eqref{eq:Relax_v_exp}, since we are only interested in the first order diffusion terms in $\eta$.
    Inserting \eqref{eq:Relax_v_exp} into \eqref{eq:Relax_psi}, we have 
    \begin{equation}
        \label{Sys:Parabolic_Relax}
        \partial_t \ppsi + \partial_x \ff(\ppsi) = \eta \partial_x \left(\left(\AA^2 - \ff^\prime(\ppsi)^2\right)\partial_x \ppsi\right).
    \end{equation}
    To obtain a diffusive approximation of the original system of equations \eqref{Sys:Hyperbolic_1D}, it has to be ensured that the diffusion term on the right hand side is non-negative. 
    This yields the so called sub-characteristic condition, that for all $\ \ppsi$ it has to hold
    \begin{equation}
        \label{eq.subchar}
        \AA^2 - \bm f^\prime(\ppsi)^2 \geq 0 \qquad \text{(positive semi definite)}.
    \end{equation} 
    Due to this restriction, we choose all entries $a_j$ of the diagonal of $\AA$ as the
    maximum absolute value over all characteristic speeds on the computational domain $\Omega$.
    Therefore we use in the following the subsequent definition 
    \begin{equation}
        \label{eq.choice.a}
        \AA^2 = \text{diag}(a^2, \dots, a^2), \quad \text{with} \quad 
        a = \underset{x\in\Omega}{\max}\underset{j=1,\dots,k}{\max}|\lambda_j(\ppsi(x,t))|.
    \end{equation}
    With this choice, the $2k$ characteristic speeds of the relaxation model \eqref{Sys:Relaxation_1D} are given by $\tilde \lambda_j = - a$ for $j = 1, \dots, k$ and $\tilde \lambda_j = a$ for $j= k+1, \dots, 2k$. 
    
    Since we consider hyperbolic equations, we can diagonalize the Jacobian by using the basis of right eigenvectors $\bm R$ and setting $\bm A$ as a constant approximation of characteristic speeds, we can write 
    \begin{equation}
        \AA^2 - \bm f^\prime(\ppsi)^2 = \bm R(\AA^2 - \bm \Lambda^2)\bm R^{-1},
    \end{equation}
    where $\bm \Lambda$ is a diagonal matrix containing the characteristic speeds $\lambda_{j}, ~j=1,\dots,k$ of the original equations \eqref{Sys:Hyperbolic_1D} given by the eigenvalues of the Jacobian $\ff^\prime(\ppsi)$.
    Therefore, $\AA^2$ as defined in \eqref{eq.choice.a} satisfies the sub-characteristic condition \eqref{eq.subchar}, see also \cite{JinXin1995}, and thus from \eqref{eq:Relax_v_exp} and \eqref{Sys:Parabolic_Relax}, we recover at leading order the original system, namely
    \begin{equation}
        \label{eq:Relax_equi_cont}
        \begin{cases}
            \vv = \ff(\ppsi), \\
            \partial_t \ppsi + \partial_x \ff(\ppsi) = 0,
        \end{cases}
    \end{equation}
    also referred to as the relaxation limit. 
    
    \subsection{Construction of the numerical scheme}
    
    In the original paper of Jin \& Xin \cite{JinXin1995}, the construction of an explicit upwind scheme was described. 
    Therein they detailed two approaches to treat the relaxation source term, namely the \textit{relaxing} and \textit{relaxed} strategy. 
    In \cite{AbbIolPup2017} a fully implicit relaxing scheme was introduced.
    A relaxing scheme is characterized by keeping the relaxation source term including the relaxation rate $\eta$ in the numerical scheme. 
    This implies that a value has to be assigned to the relaxation rate $\eta$ a priori which has to be chosen carefully to still obtain the correct relaxation limit. 
    For further details, we refer to \cite{JinLev1996} on restrictions on how to choose $\eta$. 
    Moreover, as is detailed in \cite{AbbIolPup2019}, there are additional restrictions on how to set $\eta$ in order to obtain the correct asymptotics in the low Mach limit when considering e.g. the Euler equations. 
    
    The relaxing scheme is constructed by first defining the discretization of the fluxes with an upwind scheme and then integrated in time with a backward Euler scheme. 
    This leads to the following scheme
    \begin{align}
        \label{eq:RelaxingScheme}
        \begin{split}
            &\frac{\ppsi_i^{n+1}-\ppsi_i^n}{\Delta t} + \frac{1}{\Delta x} \left(\mathcal{V}_{i+1/2}(\ppsi^{n+1},\vv^{n+1}) - \mathcal{V}_{i-1/2}(\ppsi^{n+1},\vv^{n+1})\right) = 0 \\
            &\frac{\vv_i^{n+1} - \vv_i^n}{\Delta t} + \frac{1}{\Delta x} \AA^2 \left(\mathcal{P}_{i+1/2}(\ppsi^{n+1},\vv^{n+1}) - \mathcal{P}_{i-1/2}(\ppsi^{n+1},\vv^{n+1})\right) = -\frac{1}{\eta}\left(\vv_i^{n+1} - \ff(\ppsi^{n+1}_i)\right)
        \end{split}
    \end{align}
    where $\mathcal{V}$ and $\mathcal{P}$ are the numerical fluxes associated to $\partial_x \vv$ and $\partial_x \ppsi$ respectively. 
    Since they are based on upwinding, they depend both on $\ppsi^{n+1}$ and $\vv^{n+1}$ and \eqref{eq:RelaxingScheme} is a fully coupled implicit system with $2k$ equations.
    This means the number of variables that need to be solved and updated has doubled with respect to the original problem. 
    To obtain a linear implicit system the relaxation source term was linearised using a truncated Taylor expansion. 
    This requires the knowledge of the Jacobian of the flux function which at times is difficult to compute or not available. 
    In addition, due to the coupling of variables, extensions to higher dimensions are very inefficient leading to even larger implicit systems since the number of variables increases for each added space dimension. 
    
    Therefore the aim of the numerical scheme that is presented in the next section is to reduce the numerical cost by completely eliminating the relaxation variables from the numerical scheme. 
    Furthermore we want to pass directly to the relaxation limit $\eta \to 0$ constructing a so called \textit{relaxed} implicit numerical method.
    In this way we ensure that we reach the correct relaxation limit as well as the low Mach limit without the need of fixing $\eta$ a priori, and with no need to  update the relaxation variables.

    \subsubsection{Fully implicit relaxed scheme}
    
    The numerical scheme is constructed by first deriving a time-semi discrete scheme upon which a suitable space discretization is applied. 
    The time step is given by $\Delta t = t^{n+1} - t^n$. 
    As we are interested in the limit $\eta \to 0$ to recover the original system \eqref{Sys:Hyperbolic_1D}, the relaxation source term on the right hand side of \eqref{eq:Relax_v} is stiff and is therefore discretized implicitly. 
    We apply an operator splitting and consider the following relaxation subsystem
    \begin{align}
        \begin{cases}
            \ppsi^\star &= \ppsi^n, \\[2pt]
            \vv^{\star} &= \vv^n - \frac{\Delta t}{\eta} \left(\vv^\star - \ff( \ppsi^\star)\right).
        \end{cases}
    \end{align}
    The latter equation can be rewritten and solved analytically by
    \begin{equation}
        \label{eq:Stiff_relax_impl}
        \vv^\star = \frac{\eta}{\eta + \Delta t} \vv^n + \frac{\Delta t}{\eta + \Delta t} \ff(\ppsi^\star),
    \end{equation}
    where $\ppsi^\star$ and $\vv^\star$ denote the state and relaxation variables after the relaxation process.
    Considering now the limit $\eta \to 0$, we find $\vv^\star = \ff( \ppsi^\star)$ which is consistent with the relaxation equilibrium solution \eqref{eq:Relax_equi_cont}. 
    
    Next, we discretize the homogeneous part of system \eqref{Sys:Relaxation_1D}. 
    The associated characteristic speeds are given by $\pm a_j, ~j=1,\dots,k$ and depend on the considered system of equations. 
    Especially when considering systems with fast characteristic speeds, a severe time step restriction when using an explicit scheme as done in \cite{JinXin1995} has to be enforced to guarantee stability. 
    An example for fast characteristic speeds are sound waves in the Euler equations for weakly compressible flows, or shear and longitudinal waves in hyperelastic materials, see Section \ref{sec:NonlinElast}. For further details we refer to \cite{dBraIolMil2017} and references therein. 
    Therefore, we choose a fully implicit discretization using Diagonally Implicit Runge Kutta (DIRK) methods. 
    More precisely, we apply {\em stiffly accurate} DIRK schemes, whose choice is motivated by results given in \cite{DimPar2014,ParRus2005} where a connection between the asymptotic preserving property and the structure of the DIRK scheme is investigated.
    As proven in \cite{DimPar2014} stiffly accurate DIRK schemes are also $L$ stable and are therefore in particular suited for low Mach number flows. 
    For a system of equations \eqref{Sys:Hyperbolic_1D}, an implicit Runge Kutta method based on a diagonal Butcher tableau with $s$ stages
    
    \begin{center}
        \vskip3mm
        \renewcommand{\arraystretch}{1.25}
        \begin{tabular}{c|ccc}
            $c_1$ & $\alpha_{11}$ & & \\
            $\vdots$ & $\vdots$ & $\ddots$ & \\
            $c_s$ & $\alpha_{s1}$ & $\cdots$ & $\alpha_{ss}$ \\ \hline
            & $\beta_1$ & $\cdots$ & $\beta_s$
        \end{tabular}
    \end{center}
    is given by 
    \begin{align}
        \begin{cases}
            \ppsi^{(j)} &= \ppsi^n - \Delta t \sum_{l=1}^{j} \alpha_{jl} \partial_x \ff(\ppsi^{(l)}),\\[2mm]
            \ppsi^{n+1} &= \ppsi^n - \Delta t \sum_{j=1}^{s} \beta_l \partial_x \ff(\ppsi^{(j)}). \\
        \end{cases} 
    \end{align}
    Thereby $\alpha_{jl}, \beta_l, ~j =1, \dots s, l = 1, \dots j$ denote the weights in the quadrature rule for the stages and final update respectively and $c_j$ the corresponding nodes in the considered time interval. 
    For stiffly accurate DIRK schemes the weights $\beta_l$ coincide with the weights $\alpha_{jl}$ of the last stage. 
    Applying this formalism with $s$ stages on the homogeneous part of \eqref{Sys:Relaxation_1D} leads to the following time semi-discrete scheme for $j = 1, \dots, s$ given by
    \begin{subequations}
        \label{eq:Hom_relax_impl} 
        \begin{align}
            &\begin{cases}
                \ppsi^{(j)} &= \ppsi^n - \Delta t \sum_{l=1}^{j} \alpha_{jl} \partial_x \vv^{(l)},\\[2mm]
                \vv^{(j)} &= \vv^n - \Delta t \sum_{l=1}^{j} \alpha_{jl} \AA^2 \partial_x \ppsi^{(l)},
            \end{cases} \label{eq:Hom_relax_impl_stages}\\[2mm]
            &\begin{cases}
                \ppsi^{n+1} &= \ppsi^n - \Delta t \sum_{j=1}^{s} \beta_l \partial_x \vv^{(j)}, \\[2mm]
                \vv^{n+1} &= \vv^n - \Delta t \sum_{j=1}^s \beta_l \partial_x \AA^2 \ppsi^{(j)}.
            \end{cases} \label{eq:Hom_relax_impl_update}
        \end{align} 
    \end{subequations} 
    Each stage in \eqref{eq:Hom_relax_impl_stages} forms a pairwise coupled linear system consisting of $2k$ variables that have to be solved implicitly. 
    We remark that $\bm A$ is a diagonal matrix with constant entries meaning the characteristics are frozen during one time step $(t^n,t^{n+1})$. 
    To eliminate the relaxation variables $\vv$, we insert $\vv^{(j)}$ given by the second equation of \eqref{eq:Hom_relax_impl_stages} into the first equation of \eqref{eq:Hom_relax_impl_stages} and solve for $\ppsi^{(j)}$ only. 
    The time semi-discrete scheme for the state variables $\ppsi^{(j)}$ is then given as follows
    \begin{multline}
        \label{eq:Implicit_time_semi_1}
        \ppsi^{(j)} - \Delta t^2 \alpha_{jj}^2 \AA^2 \partial_x^2 \ppsi^{(j)} = \\
        \ppsi^n - \Delta t \alpha_{jj} \partial_x \vv^n - \Delta t \sum_{l=1}^{j-1} \alpha_{jl} \partial_x \vv^{(l)} + \Delta t^2 \alpha_{jj} \sum_{l=1}^{j-1} \alpha_{jl} \AA^2 \partial_x^2 \ppsi^{(l)}.
    \end{multline} 
    At each stage, the relaxation source term is solved by using \eqref{eq:Stiff_relax_impl} leading to
    \begin{equation}
        \vv^n = \ff(\ppsi^n), \quad \vv^{(j)} = \ff(\ppsi^{(j)}).
    \end{equation}
    Thus, the obtained stage values obey an approximation to the original equations \eqref{Sys:Hyperbolic_1D}.  
    A similar structure was obtained for a second order explicit scheme introduced by Jin \& Xin in \cite{JinXin1995}.
    Using the fact that $\vv^{(l)} = \ff(\ppsi^{(l)}), ~l=1,\dots,j$ holds at relaxation equilibrium, we find 
    \begin{multline}
        \label{eq:Implicit_time_semi_2}
        \ppsi^{(j)} - \Delta t^2 \alpha_{jj}^2 \AA^2 \partial_x^2 \ppsi^{(j)} \\
        = \ppsi^n - \Delta t \alpha_{jj} \partial_x \ff(\ppsi^n) - \Delta t \sum_{l=1}^{j-1} \alpha_{jl} \partial_x \ff(\ppsi^{(l)}) + \Delta t^2 \alpha_{jj} \sum_{l=1}^{j-1} \alpha_{jl} \AA^2 \partial_x^2 \ppsi^{(l)}.
    \end{multline} 
    The stages given in \eqref{eq:Implicit_time_semi_2} form now a decoupled linear implicit system consisting only of $k$ variables, thus half of the number of variables than in \eqref{eq:Hom_relax_impl_stages}.
    As standard for implicit Runge Kutta methods, the state variables at the new time step $\ppsi^{n+1}$ are updated explicitly from the previously computed stages $\vv^{(j)}, j=1,...,s$ in \eqref{eq:Hom_relax_impl_update}. 
    Since we know that in relaxation equilibrium holds $\vv^{(j)} = \ff(\ppsi^{(j)})$, we obtain directly the solution of the relaxed system:
    \begin{equation}
        \label{eq:DIRK_final_update}
        \ppsi^{n+1} = \ppsi^n - \Delta t \sum_{j=1}^{s} \beta_j \partial_x\ff (\ppsi^{(j)}).
    \end{equation}
    As a consequence, the update of the relaxation variables $\vv^{n+1}$ in \eqref{eq:Hom_relax_impl_update} is redundant as we obtain immediately $\vv^{n+1} = \ff(\ppsi^{n+1})$ due to the relaxation process.
    Therefore, in the implementation of the numerical scheme, storing the final update for the relaxation variables $\vv^{n+1}$ can be neglected since they are given by the flux evaluation which can be computed on the fly.
    This halves the storage requirements with respect to the full relaxation scheme \cite{AbbIolPup2017}.
    Unlike for schemes directly based on the original equation \eqref{Sys:Hyperbolic_1D}, the final update \eqref{eq:DIRK_final_update} does not coincide with the last stage in \eqref{eq:Implicit_time_semi_2}. 
    This additional step is necessary to obtain an accurate description of contact waves and corrects the diffusion on the slow waves.
    
    Summarizing, the time semi-discrete scheme composed of the stages \eqref{eq:Implicit_time_semi_2} and the update \eqref{eq:DIRK_final_update} is free of relaxation variables $\vv$ and therefore we have obtained a scheme that depends only on the state variables $\ppsi$ and the storing and updating of additional variables is avoided. 
    
    To illustrate the scheme, we detail the first and second order time semi-discrete scheme based on a backward Euler scheme and second order method taken from \cite{HaiWan1991} which will be used to obtain the numerical results in Section \ref{sec:Numerics}. 
    The associated Butcher tableaux are reported in Table \ref{tab:Butcher_DIRK}.
    Note that both methods are stiffly accurate diagonal implicit Runge Kutta (DIRK) methods with a single coefficient on the diagonal (SDIRK).
    \begin{table}
        \renewcommand{\arraystretch}{1.25}
        \centering
        \begin{subtable}[b]{0.3\textwidth}
            \centering
            \begin{tabular}{c|c}
                1 & 1 \\
                \hline
                & 1 \\
            \end{tabular}
            \subcaption{Backward Euler scheme.}
        \end{subtable}\hfill
        \begin{subtable}[b]{0.65\textwidth}
            \centering
            \begin{tabular}{c|cc}
                $\gamma$ & $\gamma$ & 0 \\
                1 & $1-\gamma$ & $\gamma$ \\
                \hline 
                & $1-\gamma$ & $\gamma$ \\ 
            \end{tabular}
            \subcaption{Second order method from \cite{HaiWan1991} p. 106 with $\gamma = 1 - \frac{\sqrt{2}}{2}$.}
        \end{subtable}
        
        \caption{Butcher tableaux of the first and second order scheme.}
        \label{tab:Butcher_DIRK}
    \end{table} 
    Following the above given procedure, the first order time semi-discrete scheme with $s=1, j=1, ~\alpha_{11} = 1, ~\beta_1 = 1$ is given by 
    \begin{equation}
        \begin{cases}
            &\ppsi^{(1)} - \Delta t^2 \AA^2 \partial_x^2 \ppsi^{(1)} =
            \ppsi^n - \Delta t \partial_x \ff(\ppsi^n)	\\[2mm]
            &\ppsi^{n+1} = \ppsi^n - \Delta t \partial_x \ff(\ppsi^{(1)})
        \end{cases}
    \end{equation}
    and the second order time semi-discrete scheme with $s=2, ~\alpha_{11} = \alpha_{22} =\gamma, ~\alpha_{21} = 1-\gamma, \beta_1 = \alpha_{21}, \beta_2 = \alpha_{22}$ is then given by 
    \begin{equation}
        \begin{cases}
            &\ppsi^{(1)} - \Delta t^2 ~\alpha_{11}^2 \AA^2 \partial_x^2 \ppsi^{(1)}
            = \ppsi^n - \Delta t ~\alpha_{11} \partial_x \ff(\ppsi^n) \\[2mm]
            &\ppsi^{(2)} - \Delta t^2 ~\alpha_{11}^2 \AA^2 \partial_x^2 \ppsi^{(2)}
            = \ppsi^n - \Delta t ~\alpha_{11} \partial_x \ff(\ppsi^n) - \Delta t  \alpha_{21} \partial_x \ff(\ppsi^{(1)}) \\
            &\hskip4.8cm + ~\Delta t^2 ~\alpha_{11} \alpha_{21} \AA^2 \partial_x^2 \ppsi^{(1)}	\\[2mm]
            &\ppsi^{n+1} = \ppsi^n - \Delta t \sum_{j=1}^{2} \alpha_{2,j} \partial_x \ff(\ppsi^{(j)}).
        \end{cases}
    \end{equation}
    To further motivate the choice of L stable DIRK integrators over the one-stage well-known second order Crank-Nicholson (CN) method \cite{CraNich1947}, in the here considered context of low Mach number flows, we report the stability functions of the respective method, see also \cite{HaiWan1991} for more details.
    For a given $z \in \mathbb{C}$ the stability function $R(z)$ of the Crank-Nicholson method reads 
    \begin{equation}
        R(z) =	\frac{1 + 1/2 z}{1 - 1/2 z}
    \end{equation}
    which yields $R(z) \to -1$ in the limit $|z| \to \infty$. 
    Therefore no damping in the oscillations occurs and the asymptotics towards the incompressible limit in the low Mach number regime are not preserved. 
    The stability function of the second order DIRK method given in Table \ref{tab:Butcher_DIRK} is given as follows
    \begin{equation}
        R(z) = \frac{1 + (1- 2\gamma) z }{(1-\gamma z)^2}
    \end{equation}
    and yields $R(z) \to 0$ for $|z| \to \infty$ and is therefore L stable. 
    Moreover, since the method is also stiffly accurate, the asymptotics in the singular Mach number limit are preserved which is fundamental in the context of our applications.
    Therefore we favour the second order DIRK integrator with two stages over the Crank-Nicholson scheme with only one stage.
    
    \subsubsection{Space discretization}
    
    To obtain a fully discrete scheme, we consider the space discretization next. 
    In general one is not restricted in the choice of the space discretization keeping in mind that the numerical diffusion should be independent of the Mach number $M$. 
    Here, we choose the framework of finite volumes on a Cartesian grid.
    The computational domain $\Omega$ is divided into $N$ uniform cells $C_i = (x_{i+1/2},x_{i-1/2})$ with grid size $\Delta x$.  
    As it is standard in the finite volume setting, we consider cell averages $\ppsi_i^n$ defined at time $t^n$ by 
    \begin{equation*}
        \ppsi_i^n = \frac{1}{\Delta x} \int_{C_i} \ppsi(x,t^n) dx. 
    \end{equation*}
    For the second derivative on $\ppsi$ and the flux derivatives we apply centred differences and obtain for the stages in \eqref{eq:Implicit_time_semi_2} the following fully discrete formulation
    \begin{multline}
        \label{eq:Implicit_fully_disc_stages_centered}
        \ppsi_i^{(j)} - \Delta t^2 \alpha_{jj}^2 \AA^2 ( \ppsi^{(j)}_{i+1} - 2 \ppsi_i^{(j)} + \ppsi_{i-1}^{(j)})
        = \ppsi^n - \Delta t \alpha_{jj} (\ff_{i+1}^{n} - \ff_{i-1}^n) - \Delta t \sum_{l=1}^{j-1} \alpha_{jl} (\ff_{i+1}^{(j)} - \ff_{i-1}^{(j)})\\
        + \Delta t^2 \alpha_{jj} \sum_{l=1}^{j-1} \alpha_{jl} \AA^2 ( \ppsi^{(l)}_{i+1} - 2 \ppsi_i^{(l)} + \ppsi_{i-1}^{(l)}).
    \end{multline}
    where $\ff^n_i = \ff(\ppsi^n_i)$.
    This choice is motivated especially for low Mach number flows as it yields the correct numerical diffusion in space due to centred fluxes \cite{GuillardViozat1999,Dellacherie2010}. 
    
    To be able to apply the scheme also on compressible flows, obtaining a so called \textit{all-speed} scheme, a local Lax-Friedrichs flux can be used. 
    In compressible regimes, all characteristic speeds are of the same order and the focus lies on the resolution of all waves. 
    To obtain an all-speed scheme that can be used from compressible to weakly compressible flow regimes, we scale the diffusion by defining a function $g(M_{loc})$ based on the local Mach number $M_{loc}$ as introduced in \cite{AbbIolPup2017} to guarantee the correct numerical diffusion for all Mach regimes.
    In the following we define 
    \begin{equation*}
        g(M_{loc}) = 
        \begin{cases}
            \sin\left(\frac{\pi M_{loc}}{2}\right) &\text{for } M_{loc} \in [0,1] \\
            1 &\text{for } M_{loc} > 1
        \end{cases}.
    \end{equation*}
    Then we define the numerical fluxes as a convex combination of a centred and an local Lax-Friedrichs flux  where the convex parameter is given by $g(M_{loc}) \in [0,1]$. 
    It is given as follows
    \begin{align}
        \label{eq:Flux_convex}
        \begin{split}
            \bm f_{i+1/2} &=(1- g(M_{loc})) \frac{1}{2}\left(\bm f_i + \bm f_{i+1}\right) + g(M_{loc}) \left(\frac{1}{2}\left(\bm f_i + \bm f_{i+1}\right) - \frac{\lambda_{i+1/2}}{2}(\ppsi_{i+1} - \ppsi_i)\right)\\
            &=\frac{1}{2}\left(\bm f_i + \bm f_{i+1}\right) -  g(M_{loc})\frac{\lambda_{i+1/2}}{2}(\ppsi_{i+1} - \ppsi_i),
        \end{split}
    \end{align}
    where $\lambda_{i+1/2} = \max_{j=1,\dots,k}\max(|\lambda_j(\ppsi_{i+1})|, |\lambda_j(\ppsi_i)|)$ denotes a local approximation of the maximal characteristic speed as standard for local Lax-Friedrichs fluxes since it is applied directly on the original flux function $\ff$. 
    We emphasize, that if the local Mach number tends to zero, i.e. for low Mach number flows, we obtain a centred discretization and the numerical diffusion is independent of the Mach number.

    Since $\AA^2$ is a constant diagonal matrix, system \eqref{eq:Implicit_fully_disc_stages_centered}, with centred fluxes or with the flux \eqref{eq:Flux_convex}, consists of $k$ decoupled linear implicit equations which can be solved in parallel. 
    In addition, the number of variables to be solved in the numerical scheme corresponds to the number of state variables in the original system \eqref{Sys:Hyperbolic_1D}. 
    For the fully discrete update \eqref{eq:DIRK_final_update} is given by 
    \begin{equation}
        \ppsi_i^{n+1} = \ppsi^n - \Delta t \sum_{j=1}^{s} \beta_l \left(\ff_{i+1/2}^{(j)} - \ff_{i-1/2}^{(j)}\right).
    \end{equation}
    Note that the fully discrete scheme with the fluxes given in \eqref{eq:Flux_convex} is at most first order accurate for compressible and second order for low Mach number flows.
    
    We wish to stress the simplicity of the scheme. 
    Non-linear terms given by the fluxes $\bm f$ are evaluated explicitly and the stiffness of the low Mach case reduces to solve only linear decoupled equations. 
    Furthermore, one needs to store and update the same number of state variables as in the original problem \eqref{Sys:Hyperbolic_1D} and it is easy to implement.
    In addition, due to the decoupled linear nature of the equations that need to be solved, it is possible to use a very fine grid and still have moderate run times of the scheme. 
    
    \section{Applications}
    \label{sec:NonlinElast}
    To illustrate the properties of the above developed schemes, we apply them on two models from fluid dynamics, namely the well-known Euler equations and an extension to treat compressible materials via non-linear elasticity in a monolithic way. 
    For simplicity and sake of clarity in the development of our new numerical schemes, their derivation was given in a one dimensional framework.
    Therefore, we focus in the description of the considered models and the subsequent numerical test cases on variations along one space direction only.
    
    \subsection{The Euler equations}
    We consider the Euler equations, which are given in one dimension as follows
    \begin{align}
        \label{Sys:Euler}
        \begin{split}
            \partial_t \rho + \partial_x (\rho u) &= 0, \\
            \partial_t (\rho u) + \partial_x (\rho u^2 + p) &= 0,\\
            \partial_t E + \partial_x ((E + p) u) &=0,
        \end{split}
    \end{align} 
    where $\rho$ denotes the density, $u$ the velocity, $p$ the pressure and $E$ the total energy given by 
    \begin{equation}
        E = \rho e + \frac{1}{2} \rho u^2.
    \end{equation}
    The system is closed by choosing either the ideal gas law
    \begin{equation}
        \label{eq:EOS_idGas}
        e(\rho, p) = \frac{p}{(\gamma - 1)\rho}
    \end{equation}
    or the stiffened gas equation 
    \begin{equation}
        \label{eq:EOS_stiffGas}
        e(\rho,p) = \frac{p}{(\gamma - 1)\rho} + \frac{p_\infty}{\rho},
    \end{equation}
    where the parameter $p_\infty$ is given by the properties of the considered fluid. 
    System \eqref{Sys:Euler} can be written in conservation form \eqref{Sys:Hyperbolic_1D} setting $k=3$, $\ppsi = (\rho,\rho u,E)^T$ and $\ff(\ppsi) = (\rho u, \rho u^2 + p, u(E+p))^T$. 
    The characteristic speeds of the Euler equations consist of a material wave $\lambda_u = u$ and two acoustic waves $\lambda_\pm = u \pm c$, where $c$ denotes the sound speed. 
    For the stiffened gas equation it is given by 
    \begin{equation}
        c^2 = \gamma~\frac{p + p_\infty}{\rho}.
    \end{equation}
    Note, that by setting $p_\infty = 0$ we recover the sound speed for the ideal gas law.
    The sound speeds scale with the inverse of the \emph{acoustic Mach number} $M = |u|/c$ which is defined by the ratio between the absolute value of the fluid velocity and the sound speed of the fluid.
    Consequently, in the low Mach number regime $M \ll 1$, the acoustic wave speeds are significantly faster than the fluid velocity $\lambda_u = u$. 
    The scaling of the Euler equations with respect to the Mach number, the singular Mach number limits and the computation of characteristic speeds are extensively studied in the literature and we refer the interested reader e.g. to \cite{BoscarinoRussoScandurra2018,BouchutFranckNavoret2020,Dellacherie2010,GuillardViozat1999,KlainermanMajda1981,Klein1995,Schochet2005,ThoZenkPupKB2020,viozat1997} and references therein.
    
    \subsection{Non-linear elasticity}
    To be able to treat compressible solids in the same framework as gases and fluids described by the Euler equations \eqref{Sys:Euler}, we consider an Eulerian model for non-linear elasticity. 
    Even though we focus on a one dimensional setting, deformations of the solids are still considered in two directions. 
    Details on the derivation of the model can be found in  \cite{dBraIolMil2017,de2016cartesian,AbbIolPup2017,gorsse2014simple,plohr1988conservative,plohr1992conservative} and we shortly describe its most important features in the following.
    The equations are given by 
    \begin{align}
        \label{Sys:2D_xDir}
        \begin{split}
            \partial_t \rho + \partial_x (\rho u_1) &= 0, \\
            \partial_t (\rho u_1) + \partial_x (\rho u_1^2 - \sigma_{11}) &= 0,\\
            \partial_t (\rho u_2) + \partial_x (\rho u_1 u_2 - \sigma_{21}) &= 0,\\
            \partial_t Y_1^2 + \partial_x (u_1 Y_1^2 + u_2) &=0,\\
            \partial_t E + \partial_x ((E-\sigma_{11}) u_1 - \sigma_{21} u_2) &=0,
        \end{split}
    \end{align}
    consisting of the conservation of mass $\rho$, momentum $\rho \bm u$ and total energy $E$ which is given by the sum of the internal and kinetic energy as follows
    \begin{equation}
        \label{eq:TotE_hyper}
        E = \rho e + \frac{1}{2} \rho \Vert \bm u \Vert^2
    \end{equation}
    with $\bm{u} = (u_1, u_2)^T$.
    Further it has, in contrast to the Euler equations \eqref{Sys:Euler}, an additional equation for the deformation gradient $[\nabla Y]$ which in the one dimensional framework reduces to 
    \begin{equation}
        [\nabla Y] = 
        \begin{bmatrix}
            Y_1^1 & 0 \\
            Y_1^2 & 1 \\
        \end{bmatrix}
    \end{equation}
    with $Y_1^1 = \rho/\rho_0$, where $\rho_0$ denotes the initial density. 
    Therefore the only governing equation for the deformation gradient $[\nabla Y]$ in \eqref{Sys:2D_xDir} is given by the component $Y_1^2$. 
    
    The system \eqref{Sys:2D_xDir} is closed by considering a generalised formulation of the EOS where the gas/fluid is described by the ideal gas law \eqref{eq:EOS_idGas}/stiffened gas \eqref{eq:EOS_stiffGas} equation respectively and the compressible material as a neohookean solid, namely
    \begin{equation}
        \label{eq:EOS}
        e(\rho,p,[\nabla Y]) = \frac{p}{(\gamma - 1)\rho} + \frac{p_\infty}{\rho} + \frac{\chi}{\rho} (tr \bar B - 2).
    \end{equation}
    Therein denotes $p_\infty$ a material constant associated to the description of liquids, $\chi$ the shear modulus connected to the rigidity of the considered material and $B$ the right Cauchy-Green tensor
    \begin{equation}
        {B} = \left[\nabla Y\right]^{-1} \left[\nabla Y\right]^{-T}.
    \end{equation}
    From the latter, the modified Cauchy-Green tensor can be defined as
    \begin{equation}
        \bar{B} = \frac{B}{\text{det}\left(\left[\nabla Y \right]\right)^{-1}}.
    \end{equation} 
    Finally we obtain
    \begin{equation}
        tr \bar B = \frac{1}{\rho/\rho_0}\left(1 + (Y_1^2)^2 + (\rho/\rho_0)^2\right)
    \end{equation}
    which completes the description of the EOS \eqref{eq:EOS}.
    We want to stress that the parameters $\chi,p_\infty,\gamma$ in the EOS \eqref{eq:EOS} are determined by the given material and can be found in Table \ref{tab:Parameters_mat_tests} for the materials considered in this work. 
    
    To conclude the description of system \eqref{Sys:2D_xDir}, we can define the Cauchy stress tensor $\sigma$ appearing in the momentum and energy equations.
    It has the following non-zero components 
    \begin{equation}
        \sigma_{11} = - p + \chi \left(1 - \left(\frac{\rho}{\rho_0}\right)^2 - \left(Y_1^2\right)^2\right), \quad \sigma_{21} = -2\chi Y_1^2,
    \end{equation}
    which denote the normal and tangential stress respectively and the pressure can be computed from the EOS \eqref{eq:EOS} and the formula for the total energy density \eqref{eq:TotE_hyper}.
    
    Model \eqref{Sys:2D_xDir} can be written in conservation form \eqref{Sys:Hyperbolic_1D} with $k=5$ by setting
    \begin{equation}
        \ppsi = \begin{pmatrix}
            \rho \\ \rho u_1 \\ \rho u_2 \\ Y_1^2 \\ E
        \end{pmatrix}, \quad
        \ff(\ppsi) = 
        \begin{pmatrix}
            \displaystyle \rho u_1\\\displaystyle \rho u_1^2 + p - \chi \left(1 - \left(\frac{\rho}{\rho_0}\right)^2 - \left(Y_1^2\right)^2\right)\\\displaystyle \rho u_1 u_2  + 2\chi Y_1^2 \\u_1 Y_1^2 + u_2\\\displaystyle
            \left(E+p - \chi \left(1 - \left(\frac{\rho}{\rho_0}\right)^2 - \left(Y_1^2\right)^2\right)\right) u_1 + 2\chi Y_1^2 u_2
        \end{pmatrix}.
    \end{equation}
    
    For model \eqref{Sys:2D_xDir}, we can define two flow regimes induced by two speeds.
    We find the classical \textit{acoustic Mach number} $M$ associated with the speed of sound 
    \begin{equation}
        c(\rho,p,[\nabla Y]) = \sqrt{\gamma \ \frac{p + p_\infty}{\rho}}
    \end{equation}
    which is computed by
    \begin{equation}
        M = \frac{|u_1|}{c}.
    \end{equation}
    In addition we can define a \textit{shear Mach number} $M_\chi$ associated with an isochoric elastic speed
    \begin{equation}
        u_{iso} = \sqrt{\frac{2 \chi}{\rho}}
    \end{equation}
    which is computed by
    \begin{equation}
        M_\chi = \frac{|u_1|}{u_{iso}}.
    \end{equation}
    These Mach numbers appear in the non-dimensional formulation analysed in \cite{AbbIolPup2017} in the Cauchy tensor $\sigma$
    \begin{equation}
        \sigma_{11} = - \frac{p}{M^2} + \chi \frac{\left(1 - \left(\frac{\rho}{\rho_0}\right)^2 - \left(Y_1^2\right)^2\right)}{2 M_\chi^2}, \quad \sigma_{21} = -\chi \frac{Y_1^2}{M_\chi^2}
    \end{equation}
    and in the internal energy
    \begin{equation}
        e(\rho,p) = \frac{p}{M^2 \rho (\gamma - 1)} + \frac{p_\infty}{M^2\rho} + \frac{\chi}{2 M_\chi^2\rho} (tr \bar B - 2).
    \end{equation}
    This reflects also on the characteristic speeds, derived and analyzed in \cite{dBraIolMil2017,de2016cartesian,AbbIolPup2017,gorsse2014simple,plohr1988conservative,plohr1992conservative}, which are given, considering the parameters $M$ and $M_\chi$, by two longitudinal waves 
    \begin{equation}
        \lambda_{1,5} = u_1 \pm \sqrt{\frac{c^2}{2 M^2} + \frac{\chi}{2 \rho M_\chi^2} \left(\alpha + 1\right) + \frac{1}{\rho} \sqrt{\left(\frac{\rho c^2}{2 M^2} + \frac{\chi}{2 M_\chi^2} \left(\alpha - 1\right)\right)^2 + \frac{\chi^2}{M_\chi^2} \left(Y_1^2\right)^2}}
    \end{equation}
    where $\alpha = \left(\frac{\rho}{\rho_0}\right)^2 + \left(Y_1^2\right)^2$,
    two shear waves
    \begin{equation}
        \lambda_{2,4}  = u_1 \pm \sqrt{\frac{c^2}{2 M^2} + \frac{\chi}{2 \rho M_\chi^2} \left(\alpha + 1\right) - \frac{1}{\rho} \sqrt{\left(\frac{\rho c^2}{2 M^2} + \frac{\chi}{2 M_\chi^2} \left(\alpha - 1\right)\right)^2 + \frac{\chi^2}{M_\chi^2} \left(Y_1^2\right)^2}}
    \end{equation}
    and one material wave given by the flow velocity $\lambda_3 = u_1$. 
    For details on the Jacobian $\nabla_\ppsi \ff(\ppsi)$ and the calculation of the characteristic speeds, see e.g. \cite{AbbIolPup2017,AbbIolPup2019,dBraIolMil2017,de2016cartesian}.
    Note that by setting the shear modulus, connected to compressible solids, as $\chi = 0$, we recover the wave speeds of the classic Euler equations \eqref{Sys:Euler}, where the longitudinal waves reduce to acoustic waves and the shear waves are not present. 
    They collapse to the material wave $\lambda_{2,4} = u_1$.
    For the Eulerian model of non-linear elasticity \eqref{Sys:2D_xDir}, we can identify two different low Mach number limits given by the
    \begin{enumerate}
        \item \textit{acoustic and shear low Mach number regime} characterized by $M \ll 1$ and $M_\chi \ll 1$. 
        Both pressure and deformation gradient cause the stiffness in the equations. 
        Thus longitudinal and shear waves are significantly faster than the material wave. 
        Especially $\mathcal{O}(M) \simeq \mathcal{O}(M_\chi)$ means $c \simeq u_{iso}$ and $p + p_\infty \simeq \chi$, see Test 4 in Section \ref{sec:SimMatWaves}. 
        \item \textit{acoustic low Mach number regime} characterized by $M \ll 1$ and $M \ll M_\chi$. 
        The stiffness in the equations stems solely from the pressure resulting in $p + p_\infty \gg \chi$ and $c \gg |u_1|$ while $c \gg u_{iso}$. This means longitudinal waves are significantly faster than the shear and material waves, see Test 5 in Section \ref{sec:SimMatWaves}.
    \end{enumerate}
    
    \section{Numerical results}
    \label{sec:Numerics}
    In this section we validate the implicit first (IM1) and second order (IM2) implicit relaxed scheme at all speeds with respect to the Mach number regime applied to the Eulerian model of nonlinear elasticty \eqref{Sys:2D_xDir}.
    We compare the numerical results against an explicit local Lax Friedrichs (LLF1) scheme for which a CFL condition on the time step $\Delta t$ has to be satisfied. 
    We define an acoustic time step associated with the CFL condition
    \begin{equation}
        \label{eq:CFL_cond}
        \Delta t \leq \nu_{ac} \frac{\Delta x}{| u_1 + c|}, \quad 	\Delta t \leq \nu_{ac} \frac{\Delta x}{|\lambda_1|}
    \end{equation} 
    for gases and hyperelastic solids respectively.
    Analogously we define a material CFL condition oriented towards the flow velocity $u_1$ given by
    \begin{equation}
        \Delta t \leq \nu_{mat} \frac{\Delta x}{|u_1|}.
    \end{equation} 
    Therein $\nu_{ac}$ and $\nu_{mat}$ denote the CFL coefficient for the acoustic and material time step respectively.
    The matrix $\bm A$ with the relaxation speeds is constructed as explained in Section \ref{sec:JXrelax} and is given by $\bm A = \text{diag}(a, \dots, a)$
    %\begin{equation} \bm A =
    %\begin{pmatrix}
    %a_1 &  &&\\
    %& a_2 & &\\
    %&& a_3 &  \\
    %&&& a_4 & \\
    %&&&& a_5 \\
    %\end{pmatrix}
    %\end{equation}
    with $a = \underset{j=1,\dots,k}{\max} ~~\underset{i=1,\dots,N}{\max} ~~|\lambda_j(\ppsi_i^n)|$.
    We first perform steady test cases on the Euler equations ($\chi = 0$) by considering flows in a nozzle in different Mach number regimes. 
    Next, we present results on the propagation of material waves in gases and hyperelastic solids. 
    
    We would like to remark, that the same code with the same framework was used for {\em all} test cases presented here to ensure the consistency of our results.
    The structure of the numerical scheme presented in Section \ref{sec:NumScheme} is already quite simple from an implementation point of view, that adopting a simpler code for the Euler equations is not necessary, since the implicit systems are decoupled and can be solved independently from each other. 
    The difference between Euler equations \eqref{Sys:Euler} and the non-linear elasticity model \eqref{Sys:2D_xDir} merely consists in the number of implicit systems that need to be solved (3 for Euler and 5 for non-linear elasticity in one dimension) but their complexity remains the same independently of the considered model.

    \subsection{Laval nozzle flow}
    
\begin{figure}[t!]
        \begin{subfigure}[t]{0.5\textwidth}
            \centering
            \hskip-5mm{\includegraphics[scale=0.15]{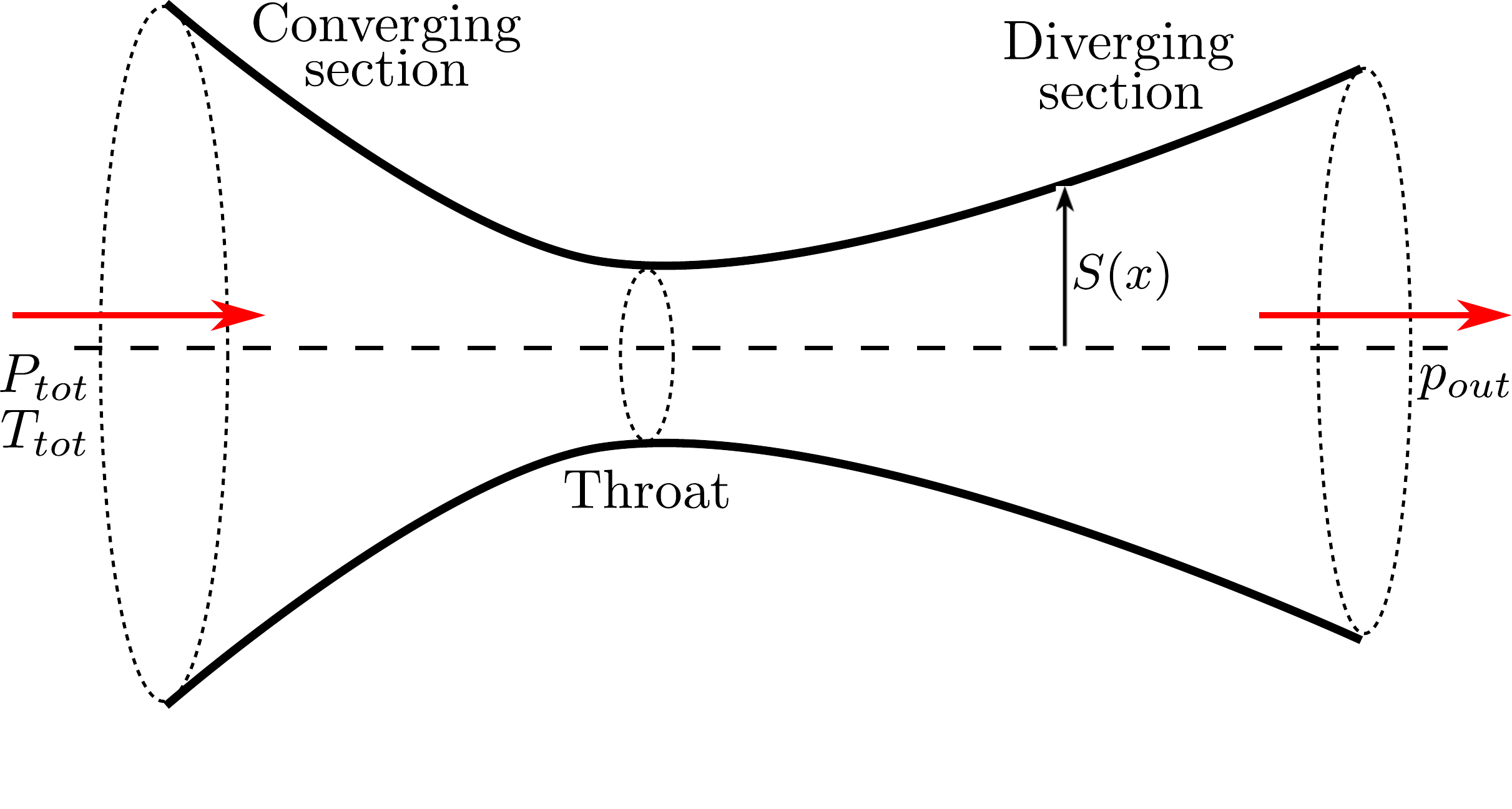} 
                \caption{Laval nozzle general sketch.}
                \label{fig:nozzle_func}}
        \end{subfigure}
        \begin{subfigure}[t]{0.5\textwidth}
            \centering
            \hskip-5mm{\includegraphics[scale=0.325]{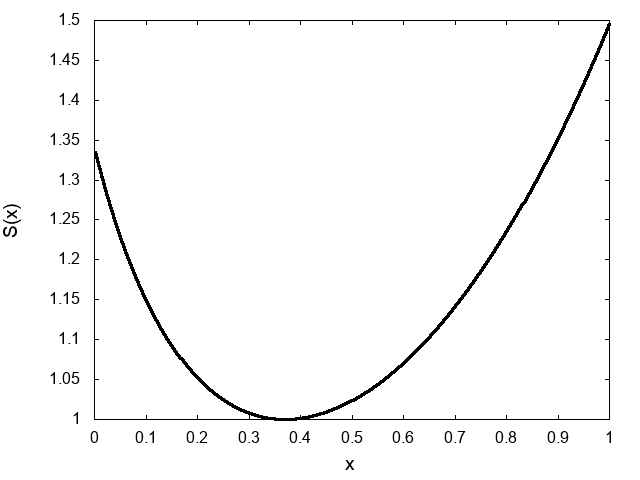}
                \caption{Geometry of the simulated nozzle.}
                \label{fig:nozzle_area}}
        \end{subfigure}
        \caption{Set-up for Laval nozzle in Test 4.1. }
        \label{fig:nozzle}
    \end{figure}

    The Laval nozzle is a converging-diverging duct. It is widely used for achieving steady supersonic flows in a variety of systems such as rocket motors and wind tunnels. The sketch of a nozzle is drawn in Figure \ref{fig:nozzle_func}. 
    
    The simplest analytic model for compressible flow in a Laval nozzle is the quasi one-dimensional duct flow approximation \cite{BenRub1971} which is a modification of the Euler equations \eqref{Sys:Euler} and is given by
    \begin{equation}\label{eq:Laval1}
        \begin{cases}
            \partial_t\left( S\rho\right) + \partial_x\left( S\rho u\right) = 0 \\
            \partial_t\left( S\rho u\right) + \partial_x\left(S\left(\rho u^2  + p\right)\right) = p\partial_xS \\
            \partial_t\left( S\rho e\right) + \partial_x\left(Su\left(\rho e+p\right)\right) = 0, \\
            \partial_t S = 0.
        \end{cases}
    \end{equation}
    The quasi one-dimensional assumption consists in taking the cross sectional area as a smooth function of the axial coordinate, $S=S\left(x\right)$ given in Figure \ref{fig:nozzle_area}. Hence, all flow variables are functions of the axial coordinate only. 
    After a few manipulations, system \eqref{eq:Laval1} can be rearranged in such a way that the Euler system \eqref{Sys:Euler} with a non linear source term is obtained
    \begin{equation}\label{eq:Laval2}
        \begin{cases}
            \partial_t\rho + \partial_x\left(\rho u\right)= -\rho u\dfrac{\partial_xS}{S} \\
            \partial_t\left(\rho u\right) +\partial_x \left(\rho u^2 + p\right) =-\rho u^2\dfrac{\partial_xS}{S}\\
            \partial_t\left(\rho e\right) + \partial_x\left(u\left(\rho e+p\right)\right) = -u\left(\rho e+p\right)\dfrac{\partial_xS}{S}\\
            \partial_t S = 0.
        \end{cases}
    \end{equation}
    Formulations \eqref{eq:Laval1}-\eqref{eq:Laval2} are equivalent and both conservative because the cross section $S\left(x\right)$ of the nozzle is a smooth function of $x$. 
    We simulate perfect gas through a Laval nozzle at different acoustic Mach number regimes.
    Thereby a steady state is reached evolving system \eqref{eq:Laval2} in time until the difference of the solution of two consecutive time steps falls below a certain tolerance, here of order $10^{-9}$. 
    More details of the set-up can be found in \cite{AvgBerIolRus2019}.
    Since it is a smooth solution, we can asses the experimental order of convergence (EOC) calculating the $L^1$ error between the numerical and steady state solution of \eqref{eq:Laval2}.
    We consider an ideal gas \eqref{eq:EOS_idGas} with $\gamma = 1.4$ and
    at the inlet of the nozzle total pressure and temperature are imposed as $P_{tot} = 1 Pa$ and $T_{tot} = 1K$, see also Figure \ref{fig:nozzle_func}. 
    The Mach number regime of the flow can be determined by choosing the outlet pressure $p_{out}$. 
    The outlet pressure and resulting Mach number regime are given in Table \ref{tab:NozzleEOC} together with the results for the scheme IM2 with $\nu_{ac} = 48$ in \eqref{eq:CFL_cond} which means that the acoustic CFL condition is largely violated.
    For low Mach number flows we achieve second order convergence, see Table \ref{tab:EOC_999,tab:EOC_99999,tab:EOC_9999999}.
    This is due to fact that in those regimes $g(M_{loc})$ is vanishingly small and the second order centred differences dominate in the numerical flux.  
    For the compressible flow in Table \ref{tab:EOC_9} with $p_{out}=0.9$, the convergence rates drop due to the significant contribution of $M_{loc}$ which gives an upwind numerical flux and the scheme is thus first order in space.
    
    It is crucial to remark that using a pure upwind scheme for the simulation of low Mach regimes yields inconsistent results as shown in \cite{AbbIolPup2017} for the exact same test case using an explicit Jin-Xin relaxation scheme. 
    For general results on the problematic of applying upwind schemes on low Mach flows see \cite{Klein1995,GuillardViozat1999,GuiMur2004,Rie2010}.
    This is mainly due to the fact that the numerical viscosity scales differently on different wave types, leading to excessive diffusion and thus inaccurate solutions.
    
    \begin{table}[t!]
        \renewcommand{\arraystretch}{1.25}
        \begin{subtable}{\textwidth}
            \begin{center}
                \begin{tabular}{ccccccc}
                    N&	$u$ & & $\rho $ & & $p$ & \\\hline\hline
                    1024& $1.553 \cdot 10^{-3}$ &---&$ 1.49 2\cdot 10^{-3}$ & --- &
                    $1.818 \cdot 10^{-3}$& --- \\ 
                    2048&$5.673\cdot 10^{-4}$&1.45& $5.163\cdot 10^{-4}$&1.53&
                    $6.277\cdot 10^{-4}$&1.53\\
                    4096& $2.443\cdot 10^{-4}$ &1.21& $2.012\cdot 10^{-4}$&1.35&
                    $2.450\cdot 10^{-4}$&1.35\\\hline
                \end{tabular}
            \end{center}
            \caption{$p_{out} = 0.9$ Pa, $M_{min} \approx 0.39, M_{max} \approx 0.72$ (compressible flow regime).}
            \label{tab:EOC_9}
        \end{subtable}
        \begin{subtable}{\textwidth}
            \begin{center}
                \begin{tabular}{ccccccc}
                    N&	$u$ & & $\rho$ & & $p$ & \\\hline\hline
                    1024&$	7.844\cdot 10^{-5}$ &---&$ 4.383\cdot 10^{-5}$ & ---& $6.086\cdot 10^{-5}$&---\\ 
                    2048&	$1.993\cdot 10^{-5}$ & 1.98& $ 1.095\cdot 10^{-5}$ &2.00&  1$.514\cdot 10^{-5}$& 2.01\\
                    4096&	$4.892\cdot 10^{-6}$ &2.02&$ 2.847\cdot 10^{-6}$ &1.94& $ 3.905\cdot 10^{-6}$&1.95 \\\hline
                \end{tabular}
            \end{center}
            \caption{$p_{out} = 0.999$ Pa, $M_{min} \approx 3.9 \cdot 10^{-2}, M_{max} \approx 7.2 \cdot 10^{-2}$.}
            \label{tab:EOC_999}
        \end{subtable}
        \begin{subtable}{\textwidth}
            \begin{center}
                \begin{tabular}{ccccccc}
                    N&	$u$ & & $\rho$ & & $p$ & \\\hline\hline
                    1024&	$8.345 \cdot 10^{-6}$ &---&  $4.660\cdot 10^{-6}$ & --- &  $6.522\cdot 10^{-6}$ & ---\\
                    2048&	$2.036\cdot 10^{-6}$ & 2.03 &  $9.945\cdot 10^{-7}$  &2.22&  $1.391\cdot 10^{-6}$ & 2.22\\
                    4096&	$4.899\cdot 10^{-7}$ & 2.05 &  $2.193\cdot 10^{-7}$ & 2.18 &  $3.070\cdot 10^{-7}$ & 2.18 \\\hline
                \end{tabular}
            \end{center}
            \caption{$p_{out} = 0.99999$ Pa, $M_{min} \approx 3.9 \cdot 10^{-3}, M_{max} \approx 7.2 \cdot 10^{-3}$.}
            \label{tab:EOC_99999}
        \end{subtable}
        \begin{subtable}{\textwidth}
            \begin{center}
                \begin{tabular}{ccccccc}
                    N&	$u$ & & $\rho$ & & $p$ & \\\hline\hline
                    1024&	$6.913\cdot 10^{-7}$ &---&  $3.163\cdot 10^{-7}$ &---&  $4.428\cdot 10^{-7}$ & --- \\
                    2048&	$1.726\cdot 10^{-7}$ &2.00&  $6.851\cdot 10^{-8}$ &2.21&  $9.591\cdot 10^{-8}$ & 2.36 \\
                    4096&	$4.076\cdot 10^{-8}$ &2.08&  $1.404\cdot 10^{-8}$ & 2.28&   $1.965\cdot 10^{-8}$ & 2.28 \\\hline
                \end{tabular}
            \end{center}
            \caption{$p_{out} = 0.9999999$ Pa, $M_{min} \approx 3.9 \cdot 10^{-4}, M_{max} \approx 7.2 \cdot 10^{-4}$.}
            \label{tab:EOC_9999999}
        \end{subtable}
        \caption{Convergence rates of scheme IM2 for Nozzle flow for different values of $p_{out}$ leading to different flow regimes with respect to the Mach number. }
        \label{tab:NozzleEOC}
    \end{table}
    
    \begin{table}[b!]
        \begin{center}
            \renewcommand{\arraystretch}{1.25}
            \begin{tabular}{cclccc}
                Test & Material & Flow regime & $\gamma$ & $p_\infty$ (Pa) & $\chi$ (Pa) \\ \hline \hline
                1 & perfect gas & $M \approx 0.9$ & 1.4 & 0 & 0\\
                2 & perfect gas & $M \approx 6 \cdot 10^{-3}$ & 1.4 & 0 & 0\\
                3 & water & $M \approx 2.5 \cdot 10^{-3}$ & 4.4 & $6.8 \cdot 10^8$ & 0\\
                4 & copper & $M \approx M_\chi \approx \mathcal{O}(10^{-3})$ & 4.22 & $3.42 \cdot 10^{10}$ & $5 \cdot 10^{10}$\\
                5 & hyperelastic solid & $M \approx 3 \cdot 10^{-3}, ~M_\chi \approx 0.15$ & 4.4 & $6.8 \cdot 10^8$ & $ 8 \cdot 10^5$ \\\hline
            \end{tabular}
            \caption{Parameters for the used materials in the tests of Section \ref{sec:SimMatWaves} and the Mach number regime of the simulated contact waves.}
            \label{tab:Parameters_mat_tests}	
        \end{center}
    \end{table}
    
    \begin{table}[h!]
        \begin{center}
            \renewcommand{\arraystretch}{1.15}
            \begin{tabular}{llllllllllll}
                Test & $\Omega$ & $x_0$ & $T_f$ & $\rho_L$ & $\rho_R$ & $u_{1,L}$ & $u_{1,r}$ & $u_{2,L}$ & $u_{2,R}$ & $p_L$ & $p_R$  \\[2pt]
                & $[m]$ & $[m] $ & $[s]$ & $\left[\frac{kg}{m^3}\right]$ & $\left[\frac{kg}{m^3}\right]$ & $\left[\frac{m}{s}\right]$ & $\left[\frac{m}{s}\right]$ & $\left[\frac{m}{s}\right]$ &  $\left[\frac{m}{s}\right]$ & $\left[\frac{kg}{m s^2}\right]$ & $\left[\frac{kg}{m s^2}\right]$\\[7pt]\hline\hline
                1 & $[0,1]$ & 0.5 &0.1644& 1 & 0.125 & 0 & 0 & 0 & 0 & 1 & 0.1\\
                2 & $[0,1]$ & 0.5 & 0.25 & 1 & 1 & 0 & $ 8 \cdot 10^{-3}$ & 0 & 0 & 0.4 & 0.399\\
                3 & $[0,1]$ & 0.5 & $10^{-4}$ & $1 \cdot 10^3$ & $1 \cdot 10^3$ & 0 & 15 & 0 & 0 & $10^8$ & $0.98 \cdot 10^8$\\
                4.1 & $[0,2]$ & 1 & $10^{-4}$ & $8.9\cdot10^3$ & $8.9\cdot10^3$ & 0 & 0 &0 & 100 & $10^9$ & $1 \cdot 10^5$\\
                4.2 & $[0,500]$ & 250 & $4\cdot 10^{-2}$ & $8.9\cdot10^3$ & $8.9\cdot10^3$ & 0 & 0 &0 & 100 & $10^9$ & $1 \cdot 10^5$\\
                5 & $[0,100]$ & 50 & 0.016 & $1 \cdot 10^3$ & $1\cdot10^3$ & 0 & 10 & 0 & 40 & $10^8$ & $ 0.98 \cdot 10^8$\\ \hline
            \end{tabular} 
            \caption{Initial condition for the material wave test cases in Section \ref{sec:SimMatWaves}.}
            \label{tab:Initial_mat_tests}
        \end{center}
    \end{table}
    \subsection{Simulation of material waves}
    \label{sec:SimMatWaves}
    In this section, we solve Riemann problems (RPs) for different materials where our interest lies in the motion and accurate capturing of material waves in different Mach regimes.
    The set-up of the test cases are taken from \cite{AbbIolPup2017}.
    % and following \cite{AbbIolPup2017}, we apply a smoothing using arc tangent function on the initial conditions.
    The parameters and initial conditions for each test case are given in Table \ref{tab:Parameters_mat_tests,tab:Initial_mat_tests}.
    For all test cases we use Neumann boundary conditions where we impose $\frac{\partial \ppsi}{\partial x} = 0$. 
    Tests 1 to 3 concern the Euler equations \eqref{Sys:Euler} whereas Tests 4 and 5 concern the model of non-linear elasticity \eqref{Sys:2D_xDir}.
    To demonstrate the effect of the final update \eqref{eq:DIRK_final_update} in the DIRK formalism on the capturing of the contact wave, we give the {\em predicted} solution given by the stage \eqref{eq:Implicit_time_semi_2} for the first order scheme. It is indicated by IM1p.
    
    \subsubsection{Perfect gas}
\begin{figure}[t!]
        \begin{center}
            \includegraphics[scale=0.397]{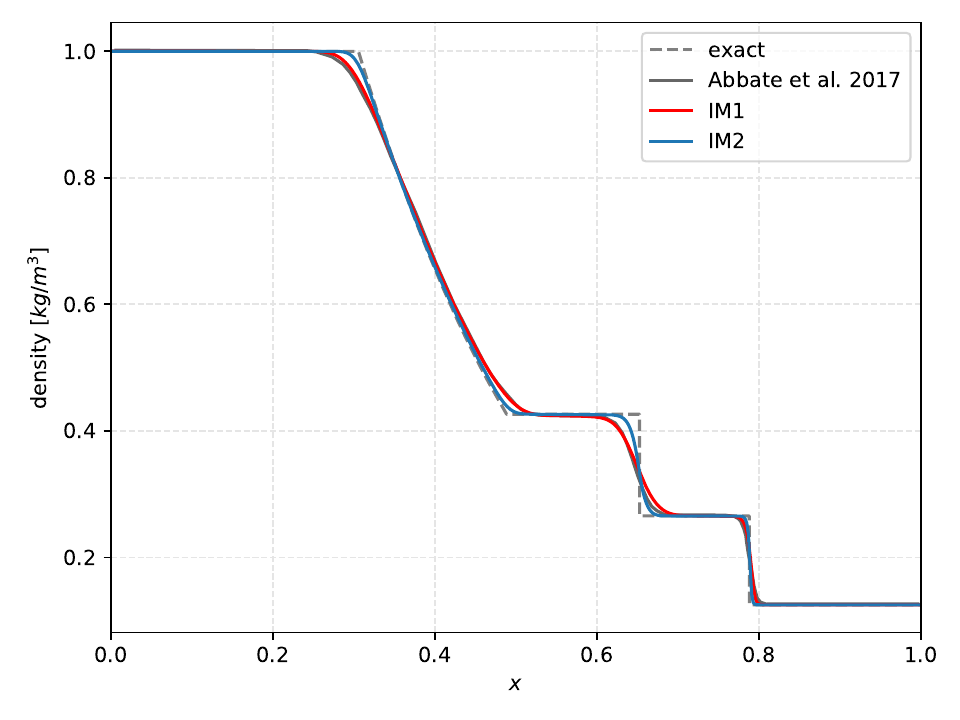}
            \includegraphics[scale=0.397]{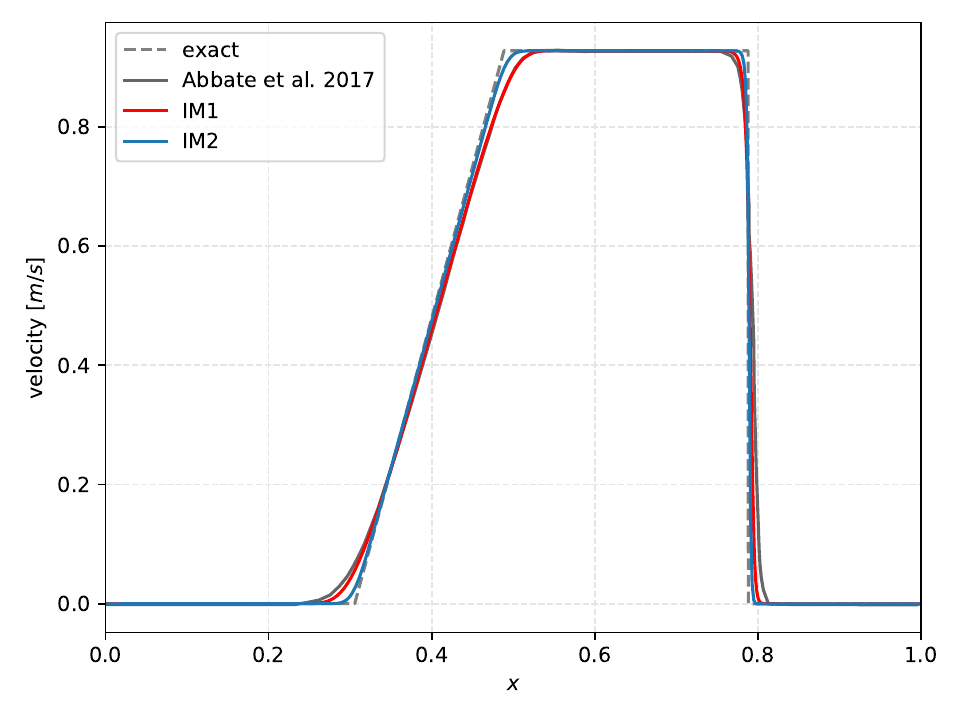}
            \includegraphics[scale=0.397]{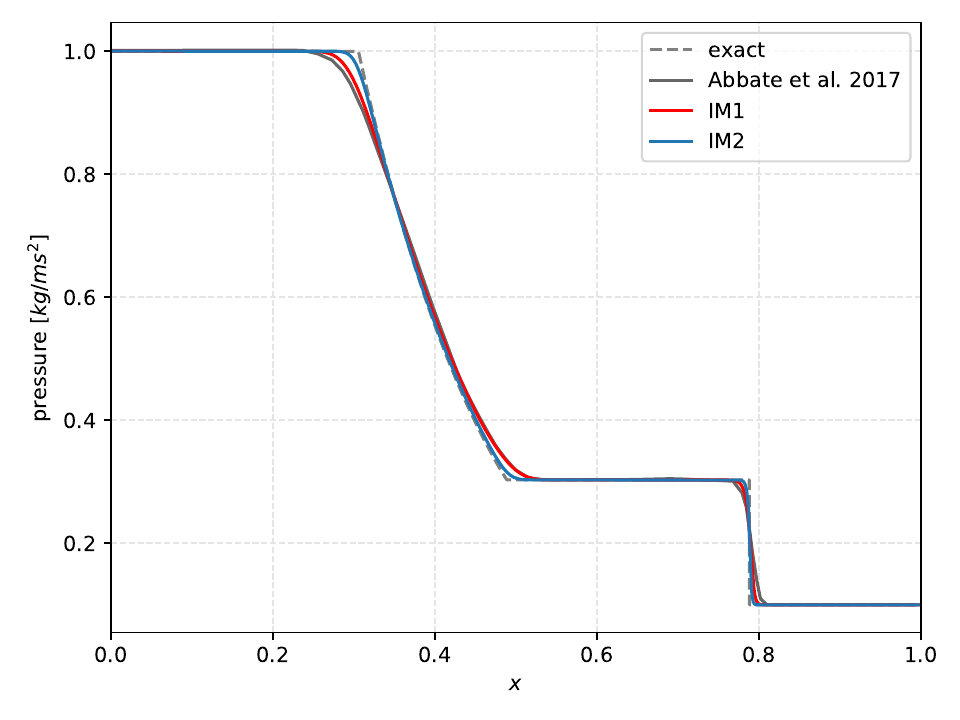}
        \end{center}
        \caption{Test 1: Sod shock tube with ideal gas on computational domain $[0,1]$ with 500 grid cells for the scheme IM2 and 1000 grid cells for IM1 and the relaxing first order scheme presented in \cite{AbbIolPup2017}.}
        \label{fig:Sod}
    \end{figure}

    \textbf{Test 1} is the Sod shock tube test case with a bi-atomic perfect gas in the compressible regime. 
    The solution of the RP consists of a rarefaction, a contact discontinuity and a shock wave. 
    Since the test is situated in the compressible regime, we apply an acoustic CFL condition with $\nu_{ac} = 0.9$. 
    For scheme IM1 we divide the computational domain in $N=1000$ grid cells to have the same discretization in space than the results of the fully coupled relaxing scheme  \eqref{eq:RelaxingScheme} presented in \cite{AbbIolPup2017} in Fig. 7. 
    For the second order scheme IM2 the computational domain is divided into $N=500$ grid cells.  
    This results in the same amount of computational time, since scheme IM2 consists of two stages. 
    Furthermore we apply a minmod reconstruction in the explicit diffusion of scheme IM2 to obtain a fully second order scheme in space and time. 
    The numerical results are given in Figure \ref{fig:Sod} and are in good agreement with the exact solution of the RP. 
    Schemes IM1 and IM2 both capture all waves accurately with the correct shock strength and speed, whereas scheme IM2 is more accurate on the contact wave than scheme IM1. 
    Moreover, our first order scheme IM1 is in good agreement with the fully coupled relaxing scheme \eqref{eq:RelaxingScheme} presented in \cite{AbbIolPup2017}.
    
\begin{figure}[t!]
        \begin{center}
            \includegraphics[scale=0.397]{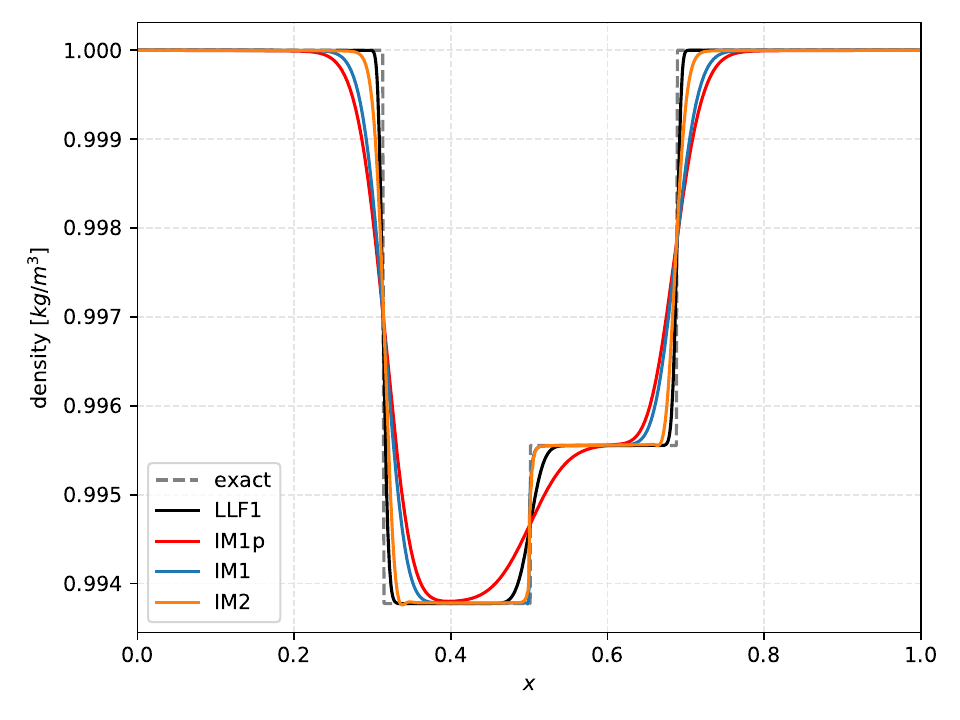}
            \includegraphics[scale=0.397]{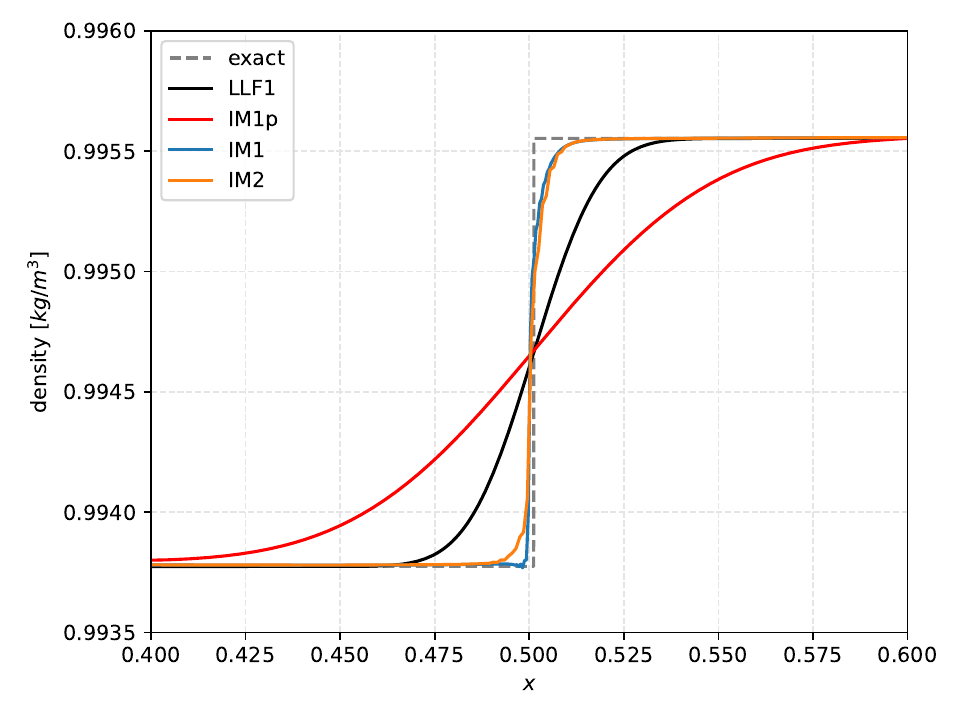}
            \includegraphics[scale=0.397]{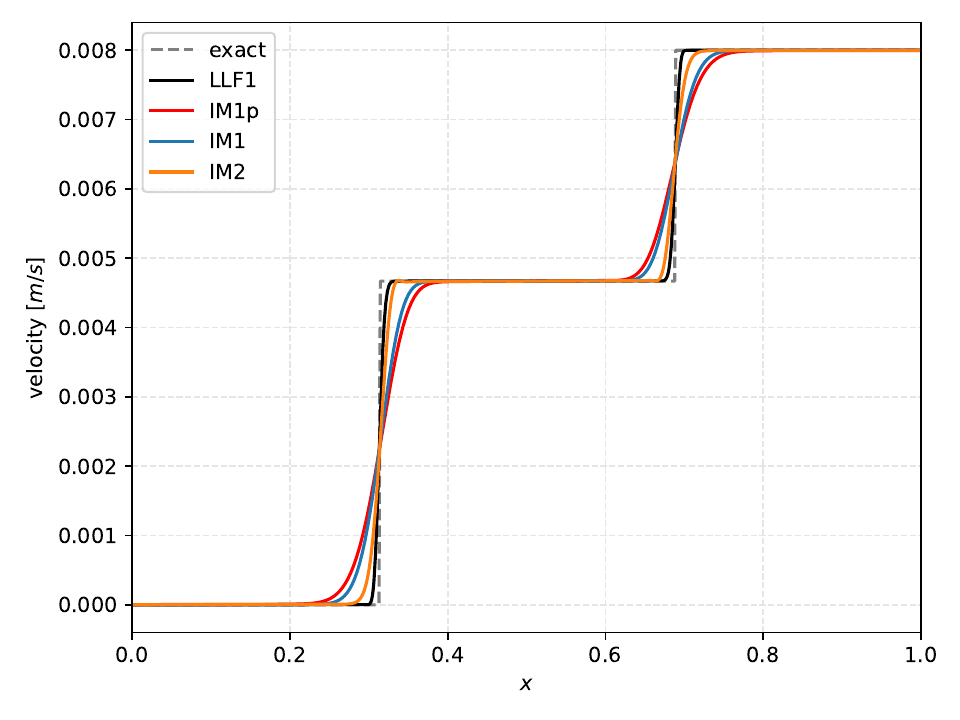}
            \includegraphics[scale=0.397]{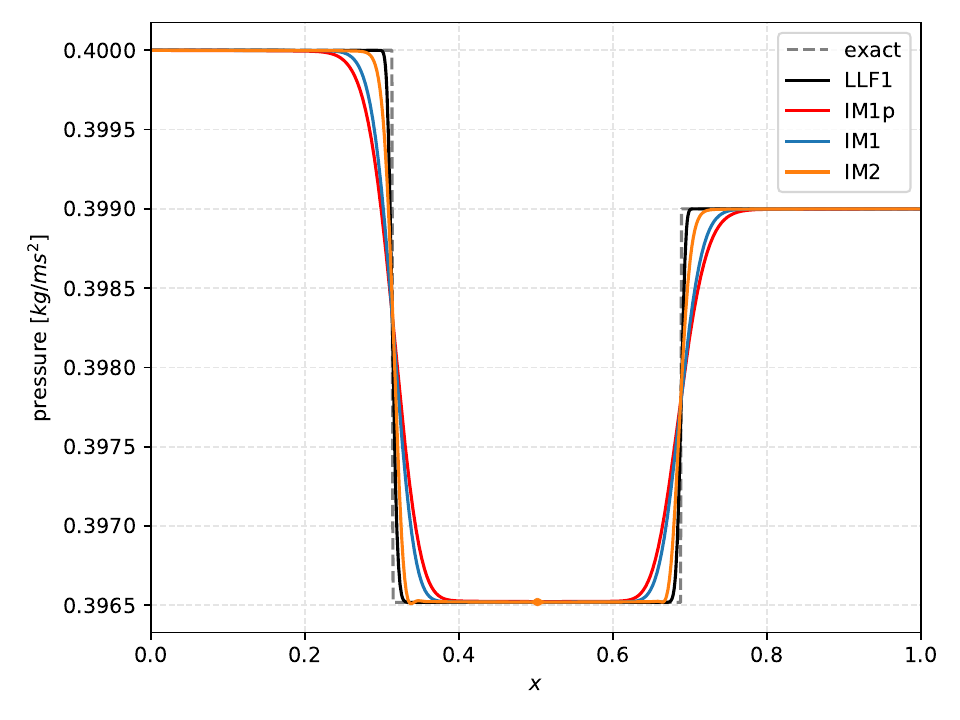}
        \end{center}
        \caption{Test 2: Low Mach tube with ideal gas on computational domain $[0,1]$ with 1000 grid cells for scheme IM2 and 2000 grid cells for the schemes IM1p, IM1 and LLF1. Top right: Zoom on the contact wave in density.}
        \label{fig:Test2}
    \end{figure}

    In \textbf{Test 2} a RP in the low Mach regime with a local Mach number $M_{loc} \approx 6 \cdot 10^{-3}$ on the material wave is considered. 
    The left and right travelling acoustic waves are significantly faster than the contact wave which travels only a few cells during the simulation. 
    We compare schemes IM1p and IM1 with $N=2000$ and scheme IM2 with $N=1000$ grid cells to an explicit first order local Lax-Friedrichs (LLF1) scheme with $N=2000$ grid cells. 
    The time steps for the LLF1 scheme are restricted by the fastest acoustic wave, i.e.  $\nu_{ac} = 0.9$ resulting in $\Delta t = 6\cdot10^{-4}$. 
    For the implicit relaxation schemes we use larger time steps given by $\Delta t = 3\cdot10^{-3}$ for scheme IM1 and $\Delta t = 6\cdot10^{-3}$ for scheme IM2. 
    The results are given in Figure \ref{fig:Test2}. 
    As expected for large time steps, scheme IM1p, using only the predicted solution $\psi^{(1)}$, is consistently diffusive on the material wave, where the solution {\em corrected} by the final update captures the material wave well.
    Scheme IM2 captures the material wave well and is more accurate on the acoustic waves than the first order schemes.
    For a comparison of the results with the fully coupled scheme \eqref{eq:RelaxingScheme} detailed in \cite{AbbIolPup2017}, see Fig. 9 in \cite{AbbIolPup2017}.
    
    To further quantify this observation, we compare the $L^1$-error of the density around the contact wave on the interval $[0.4,0.6]$ versus the needed CPU time in Table \ref{tab:Test2CPU}.
    The CPU time is obtained by averaging the computational times over 50 simulations. 
    As reference, the exact solution of the Riemann problem is used which can be obtained by procedures given in \cite{Toro2009}. 
    We see that with the LLF1 scheme, we need a very fine grid to achieve similar errors on the contact wave as produced by scheme IM2. 
    A fine grid however imposes a very strict constraint on the CFL condition leading to very small time steps and thus long CPU times. 
    Since we can choose large time steps for scheme IM2, we have smaller CPU times than for scheme LLF1 on the same grid, even though we have to solve implicit systems in scheme IM2. 
    This demonstrates the advantage of the implicit scheme IM2 over explicit scheme LLF1 for the resolution of material waves in low Mach regimes. 
    To achieve the same error given by scheme IM2, the explicit scheme LLF1 needs much finer grids and much longer CPU times.
    
    \begin{table}[htbp]
        \renewcommand{\arraystretch}{1.25}
        \begin{center}
            \begin{tabular}{lccrc}
                scheme & $\Delta x$ & $\Delta t$ & CPU time [s] & $L^1$-error \\\hline\hline
                IM2 & $10^{-3}$ & $6.00\cdot10^{-3}$ &$0.348$ & $4.20 \cdot 10^{-6}$ \\\hline
                LLF1 & $10^{-3}$ & $1.19\cdot10^{-3}$&$0.371$ & $2.00 \cdot 10^{-5}$ \\
                LLF1 & $10^{-4}$ & $1.19\cdot10^{-4}$ &$17.609$ & $7.22 \cdot 10^{-6}$ \\
                LLF1 & $10^{-5}$ & $1.19\cdot10^{-5}$ &$1587.385$ & $3.32 \cdot 10^{-6}$ \\\hline
            \end{tabular}
        \end{center}
        \caption{Test 2: $L^1$-error in $\rho$ vs. CPU time on the contact wave on the interval $[0.4,0.6]$. }
        \label{tab:Test2CPU}
    \end{table}
    
    \subsubsection{Stiffened gas}
\begin{figure}[h!]
        \begin{center}
            \includegraphics[scale=0.397]{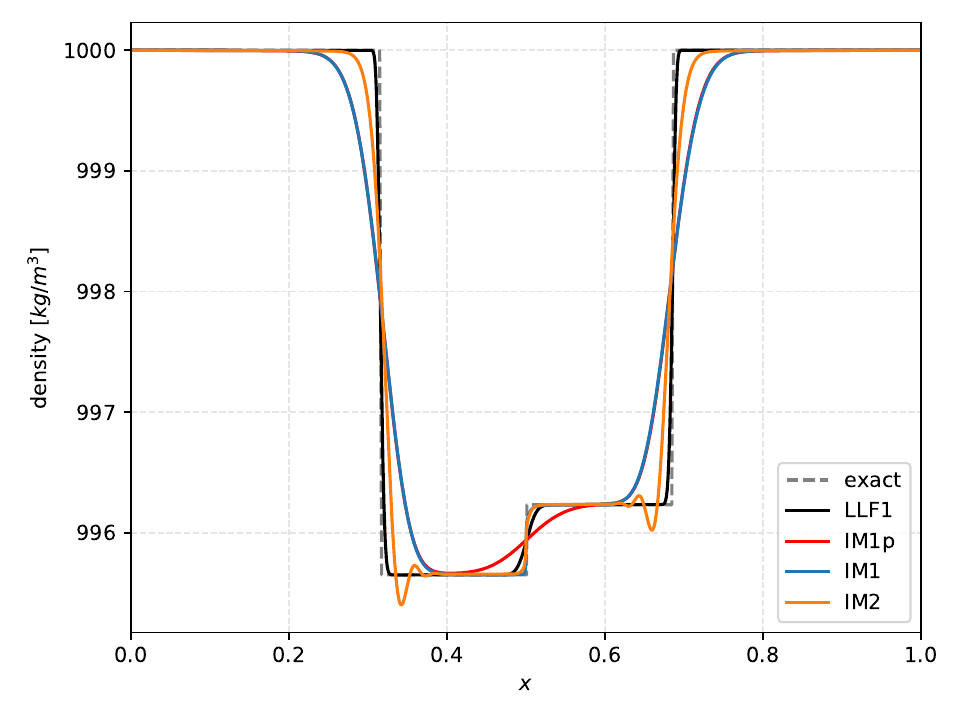}
            \includegraphics[scale=0.397]{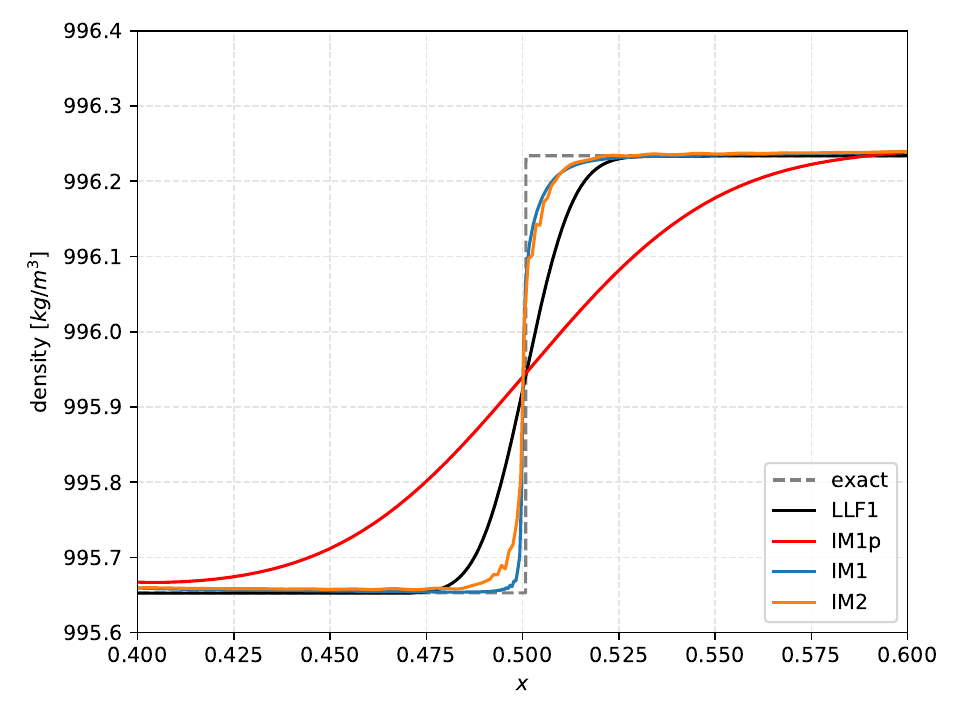}
            \includegraphics[scale=0.397]{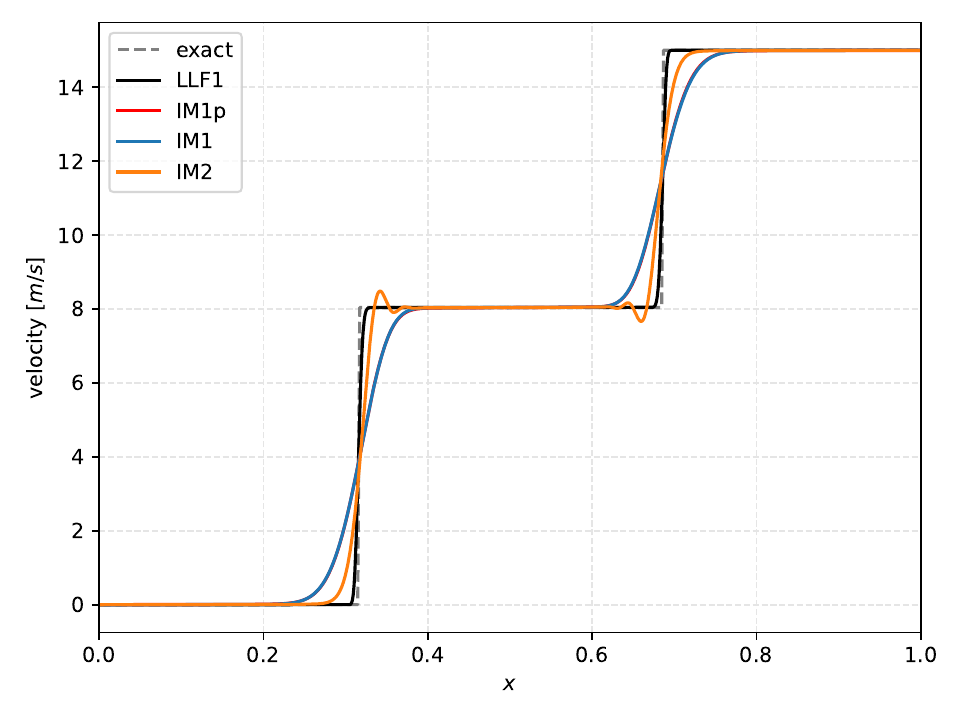}
            \includegraphics[scale=0.397]{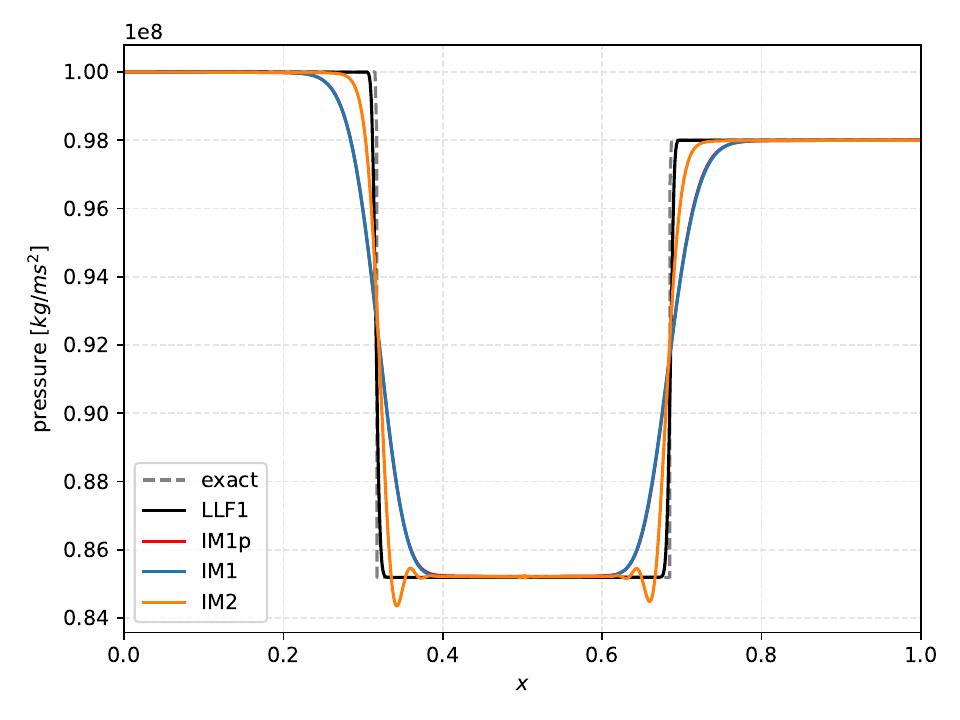}
        \end{center}
        \caption{Test 3: Tube with water on the computational domain $[0,1]$ with 1000 grid cells for scheme IM2 and 2000 grid cells for the schemes IM1p, IM1 and LLF1. Top right: Zoom on the contact wave in density.}
        \label{fig:Test3}
    \end{figure}

    \textbf{Test 3} concerns water flow in a pipe with a small pressure jump. 
    The local Mach number on the contact wave is given by $M_{loc} \approx 2.5\cdot10^{-2}$ and the set up is analogous to Test 2 for an ideal gas. 
    The time step for the LLF1 scheme must be oriented towards the fastest acoustic wave with $\nu_{ac} = 0.9$ which results in a time step of $\Delta t = 2.4\cdot10^{-7}$. 
    For the implicit scheme we can choose larger time steps given by $\Delta t = 2.15\cdot10^{-6}$ for schemes IM1p and IM1 and $\Delta t = 4.30\cdot10^{-6}$ for scheme IM2 which corresponds to $\nu_{mat} = 0.06$ or respectively $\nu_{ac} = 8.1$ in the CFL condition.
    Since the sound speeds are faster in water than for an ideal gas, the time steps are significantly smaller compared to Test 2.  
    The numerical results are given in Figure \ref{fig:Test3}. 
    Scheme As in the previous test case, IM1p is very diffusive on the acoustic as well as on the material wave,  whereas schemes IM1 and IM2  are sharp on the material wave and resolve it more accurately than the scheme LLF1. 
    However, scheme IM2 is inaccurate on the negligible acoustic waves where small oscillations can be observed. 
    Their appearance is local and does not impair the results on the contact wave which are the focus of the simulation.
    For a comparison of the results with the fully coupled scheme \eqref{eq:RelaxingScheme} given in \cite{AbbIolPup2017}, see Fig. 10 in \cite{AbbIolPup2017}.
    
    \subsubsection{Hyperelastic solid}
    The next two test cases concern deformations of hyperelastic solids using the model of non-linear elasticity \eqref{Sys:2D_xDir}.
    The tests illustrate the two different low Mach regimes described at the end of Section \ref{sec:NonlinElast}.
    
\begin{figure}[t!]
        \begin{center}
            \includegraphics[scale=0.397]{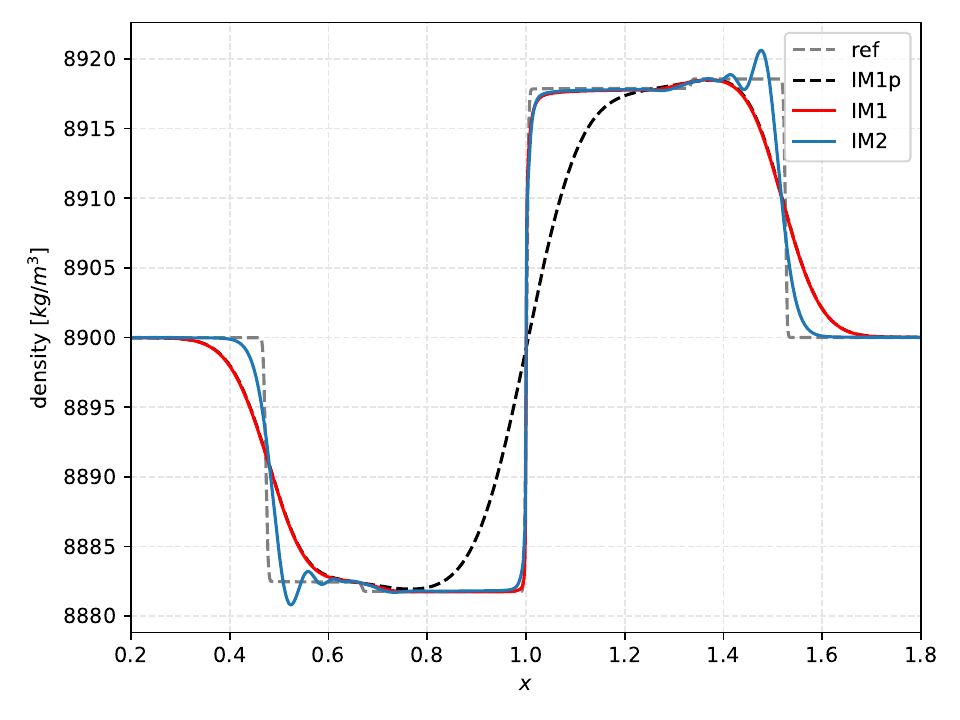}
            \includegraphics[scale=0.397]{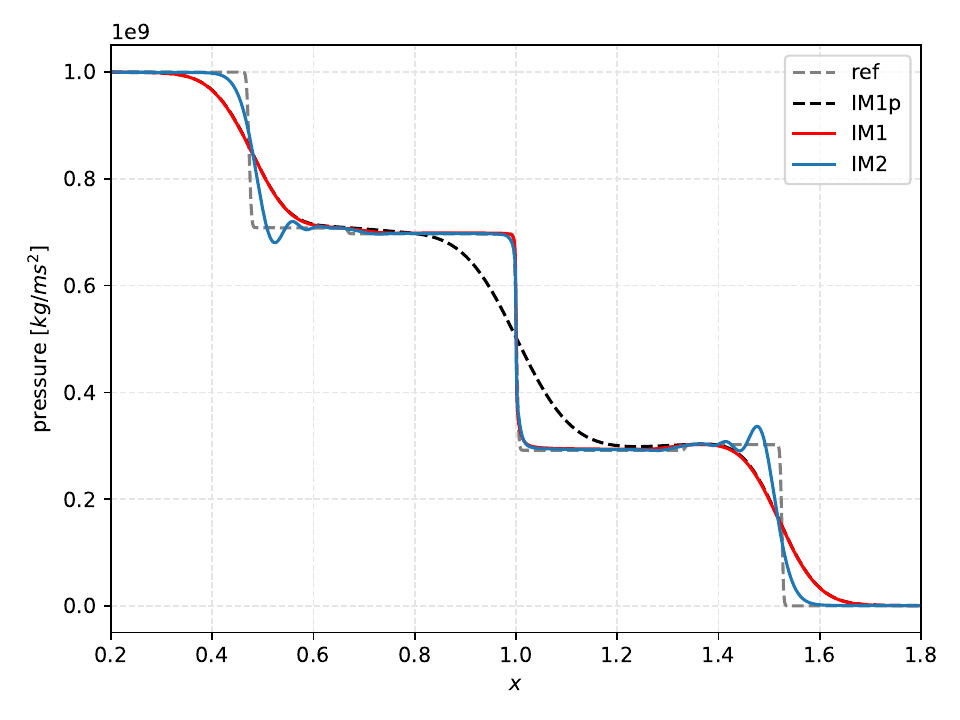}
            \includegraphics[scale=0.397]{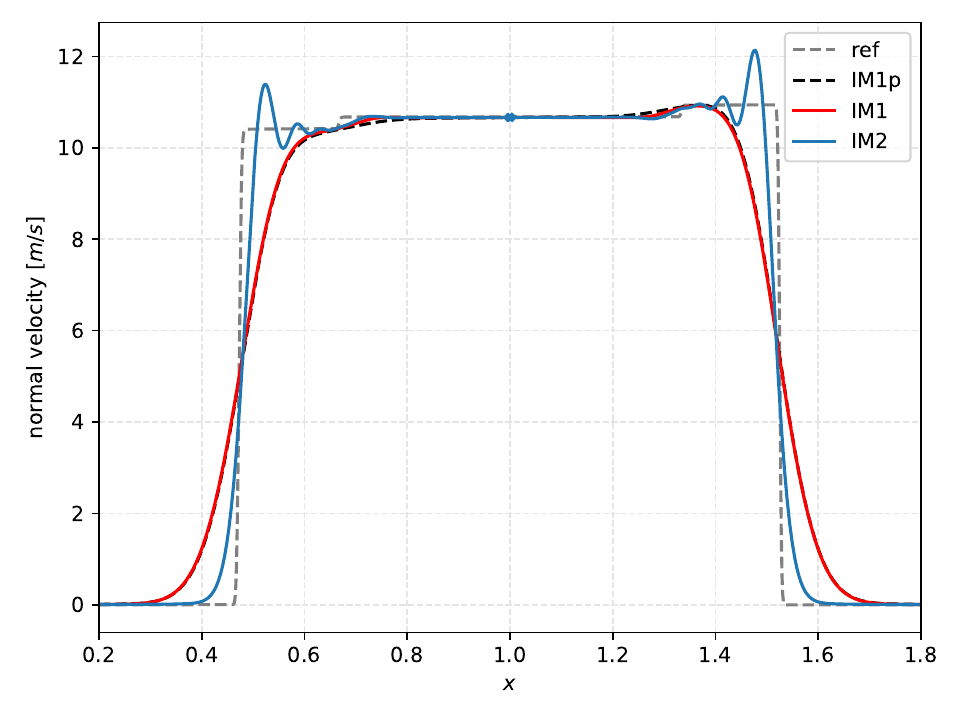}
            \includegraphics[scale=0.397]{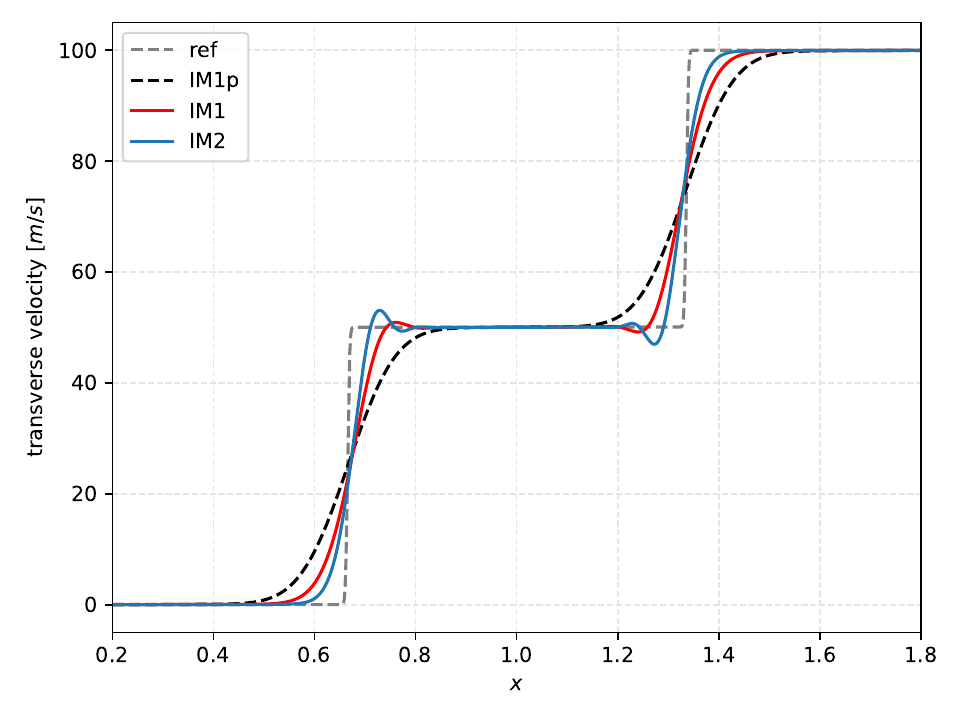}
            \includegraphics[scale=0.397]{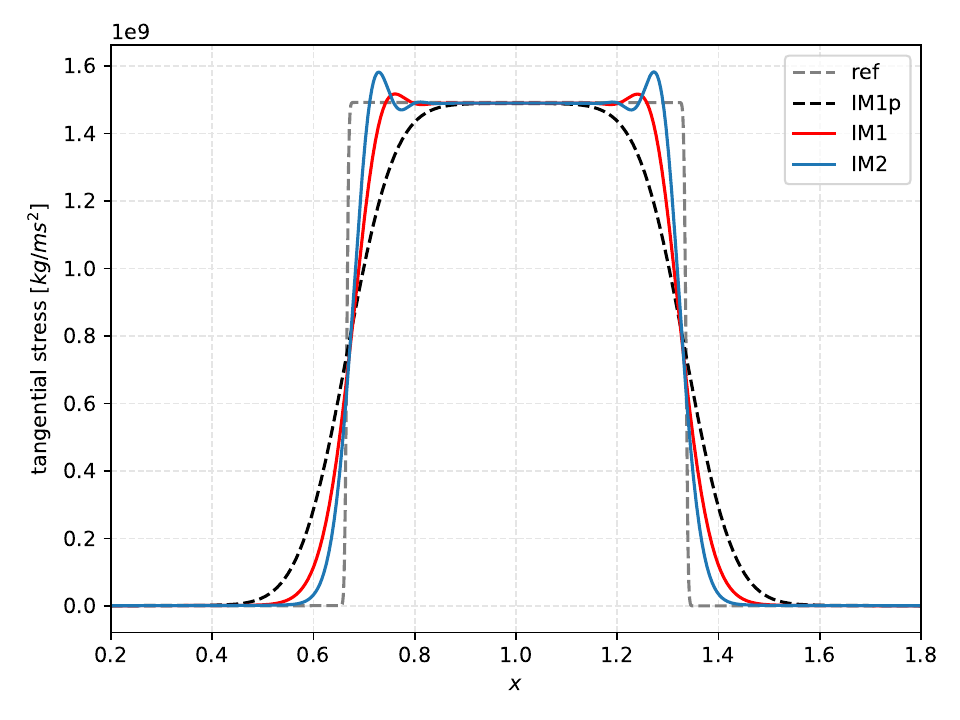}
            \includegraphics[scale=0.397]{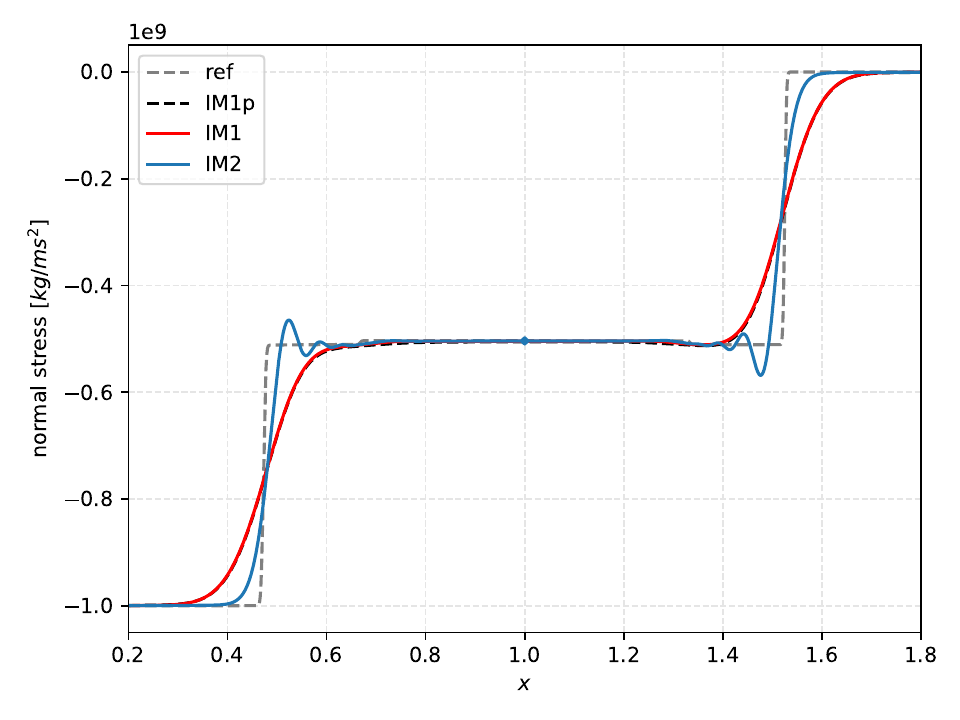}
        \end{center}
        \caption{Test 4.1: Tube with copper on the computational domain $[0,2]$ for short times with 2000 grid cells for scheme IM2 and 4000 grid cells for scheme IM1. The reference solution is computed with a second order explicit LLF scheme with $10^5$ grid cells.}
        \label{fig:Test41}
    \end{figure}

    \textbf{Test 4.1} simulates a deformation of a two meter pipe filled with copper. 
    Initially the copper is at rest with normal velocity $u_1 = 0$.
    A higher pressure is applied on the left part of the pipe. 
    A non-zero tangential velocity $u_2$ is imposed on the right part of the pipe which leads to a RP consisting of 5 waves. 
    Since $p_\infty$ and $\chi$, given in Table \ref{tab:Parameters_mat_tests}, are of the same magnitude, the longitudinal and shear waves are both significantly faster than the material wave which is situated in a flow regime of $\mathcal{O}(10^{-3})$. 
    This corresponds to a low acoustic and low shear Mach number regime.
    For the schemes IM1p and IM1 we use $N=4000$ grid cells and for scheme IM2 $N=2000$ grid cells on the domain $[0,2]$.   
    The time step for an explicit scheme imposed by the fastest wave with a CFL condition of $\nu_{ac} = 0.9$ results in $\Delta t = 3.15\cdot10^{-8}$. 
    However, for the implicit relaxation schemes larger time steps can be used given by $\Delta t = 1.25\cdot10^{-6}$ for schemes IM1p and IM1 and $\Delta t = 2.5\cdot10^{-6}$ for scheme IM2 which corresponds to $\nu_{mat} = 0.25$ or respectively $\nu_{ac} = 35.7$ in the CFL condition.
    The numerical results are given in Figure \ref{fig:Test41}. 
    The reference solution is obtained with an explicit second order scheme using local Lax-Friedrichs fluxes and a strong stability preserving Runge Kutta method (SSPRK2) on a fine grid with $10^5$ cells. It captures all waves accurately at the cost of being restricted to very small time steps. 
    The predicted solution by scheme IM1p is too diffusive to capture the material wave whereas the schemes IM1 and IM2 detect the contact wave very accurately even though time steps 36 times larger than the acoustic one are used. 
    However, we observe local oscillations on the fast longitudinal and shear waves for scheme IM2.
    We want to stress that the focus of the simulation lies on a sharp resolution of the material wave which is unaffected by the oscillations.
    
\begin{figure}[t!]
        \begin{center}
            \includegraphics[scale=0.397]{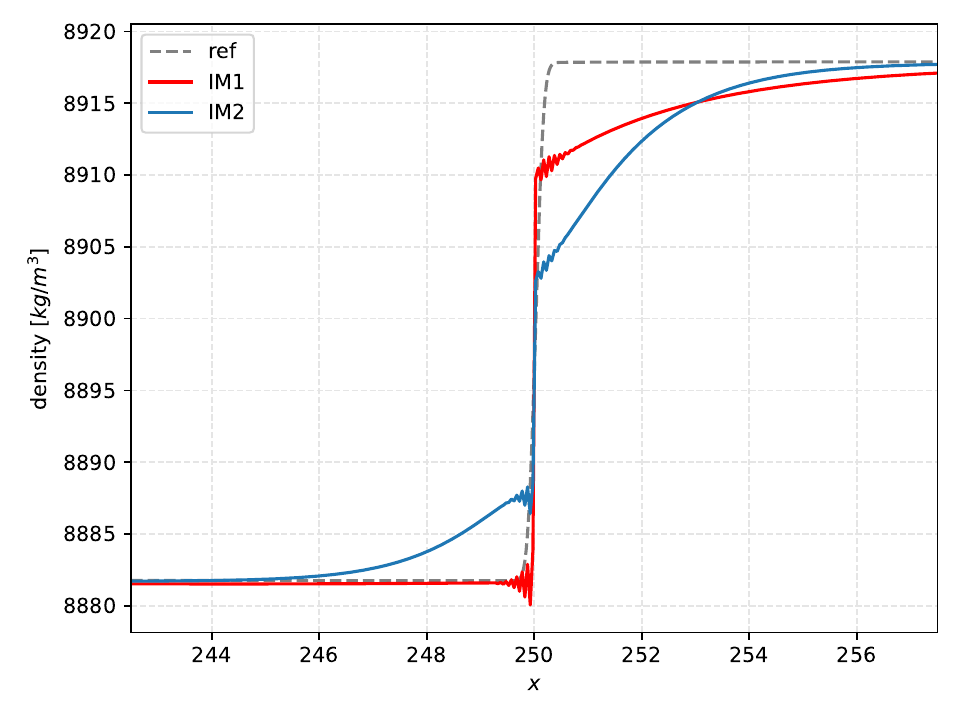}
            \includegraphics[scale=0.397]{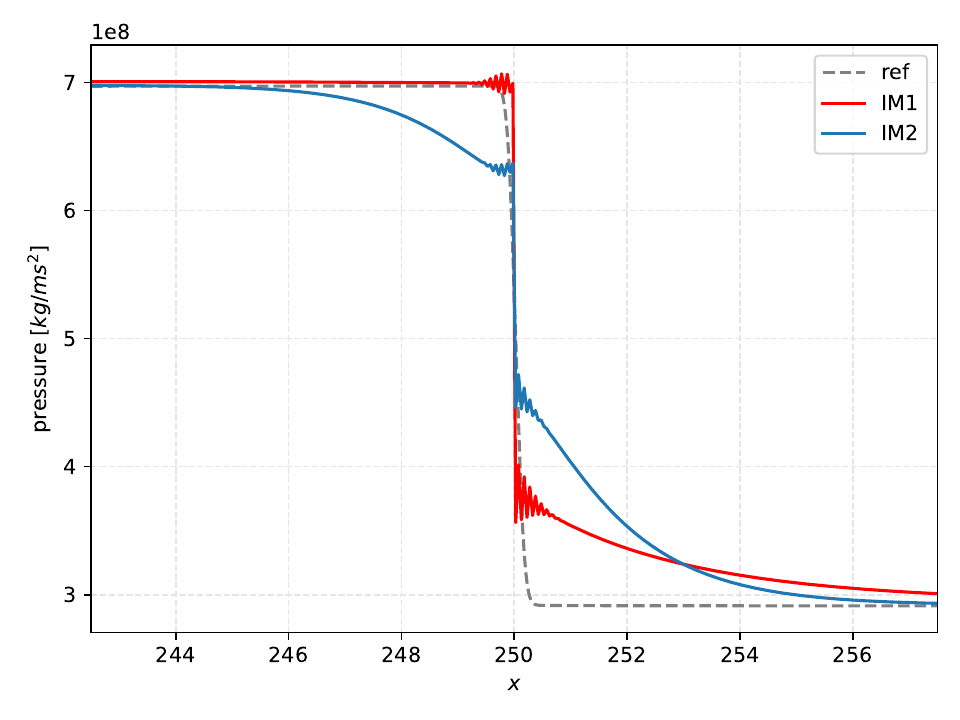}
        \end{center}
        \caption{Test 4.2: Tube with copper on the computational domain $[0,500]$ for longer times with $\Delta x = 5\cdot10^{-2}$ for the schemes IM1 and IM2. Focus on the material waves in density and pressure on the subdomain $[243,257]$ around the initial discontinuity $x_0 = 250$.  }
        \label{fig:Test42}
    \end{figure}

    To asses this further, we run Test 4.1 over a longer time period. 
    The set-up is given in \textbf{Test 4.2} in Table \ref{tab:Initial_mat_tests}. 
    The length of the pipe of 500m is chosen such that all waves are still contained in the computational domain and no boundary effects arise. 
    We run all schemes with 10000 grid cells which amounts to $\Delta x = 5\cdot10^{-2}$. 
    For the implicit schemes IM1, IM2 we impose a large time step of $\Delta t = 3.8 \cdot 10^{-4}$ about 200 times larger than the acoustic constraint with $\nu_{ac} = 0.9$ leading to a time step of $\Delta t = 8.5\cdot 10^{-6}$.
    In Figure \ref{fig:Test42}, the material wave in the density and pressure for schemes IM1 and IM2 are given. 
    Both schemes capture the position of the contact wave accurately even over long simulation times.
    As reference solution, the results of the contact wave with the explicit scheme SSPRK2 of Test 4.1 is used.
    This is motivated by the fact, that the contact wave travels at most one cell for the chosen grid in the simulation with the schemes IM1 and IM2 and moreover, running the whole simulation with the explicit scheme would have diffused the material waves to an extend that it would not be feasible as reference solution.
    Therefore the material wave of the reference solution is slightly smoothed out. 
    The numerical results on the material wave in density and pressure for Schemes IM1 and IM2 validate that the oscillations on the acoustic waves observed in Figure \ref{fig:Test41} do not affect the material waves which is the main interest of the simulation.
    For a comparison of the results with the fully coupled scheme \eqref{eq:RelaxingScheme} given in \cite{AbbIolPup2017}, see Fig. 11 to 13 therein.
    We would like to remark that due to the decoupled structure of the new numerical scheme, a considerable reduction of computational overhead is obtained for a fixed number of cells. 
    We observe that we measure a comparable CPU time using a space discretization of 10000 cells in the new scheme, compared to the CPU time needed in \cite{AbbIolPup2017} with 4000 cells on the domain [0, 500]. However the results are drastically improved, see Figure \ref{fig:Test42}, yielding a much more accurate capturing of the material waves. 
    This demonstrates the advantage of the new scheme especially with respect to large domains and long time simulations.

\begin{figure}[t!]
        \begin{center}
            \includegraphics[scale=0.38]{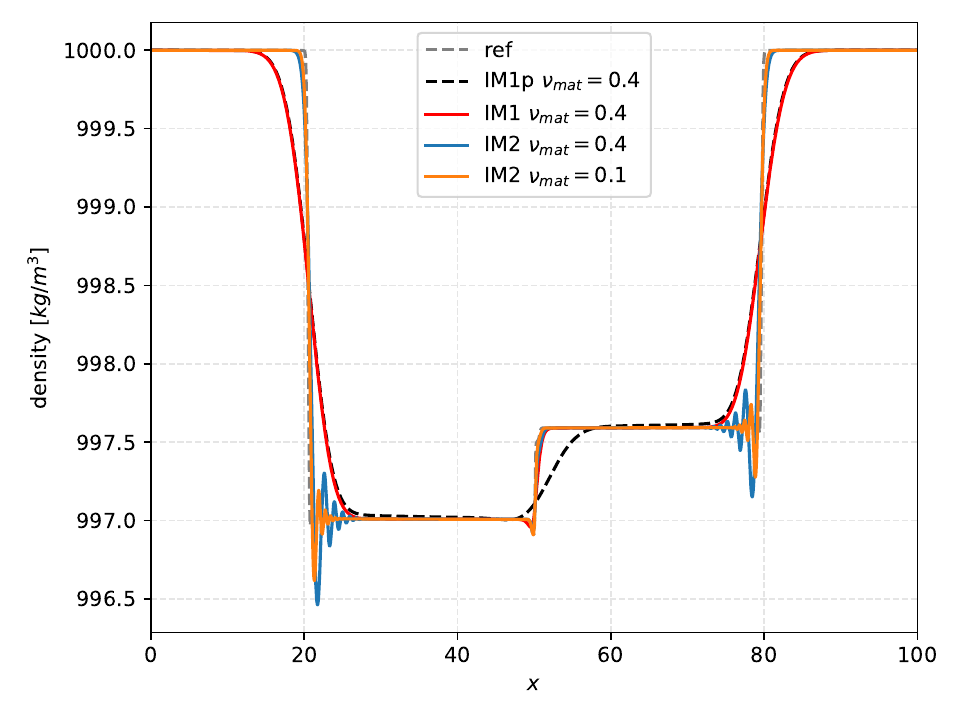}
            \includegraphics[scale=0.38]{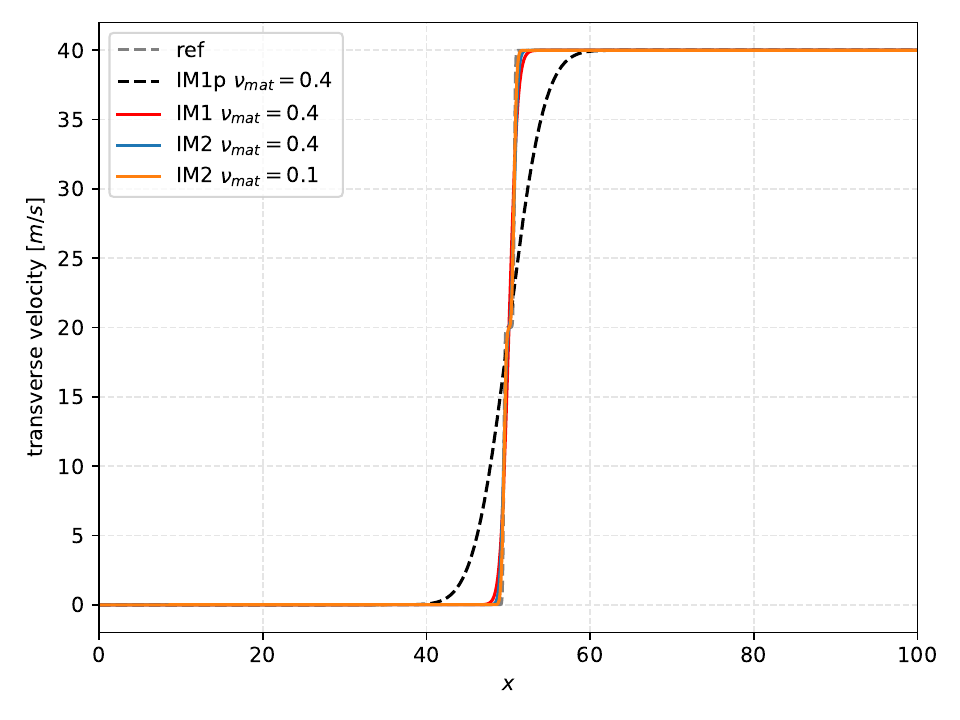}
            \includegraphics[scale=0.38]{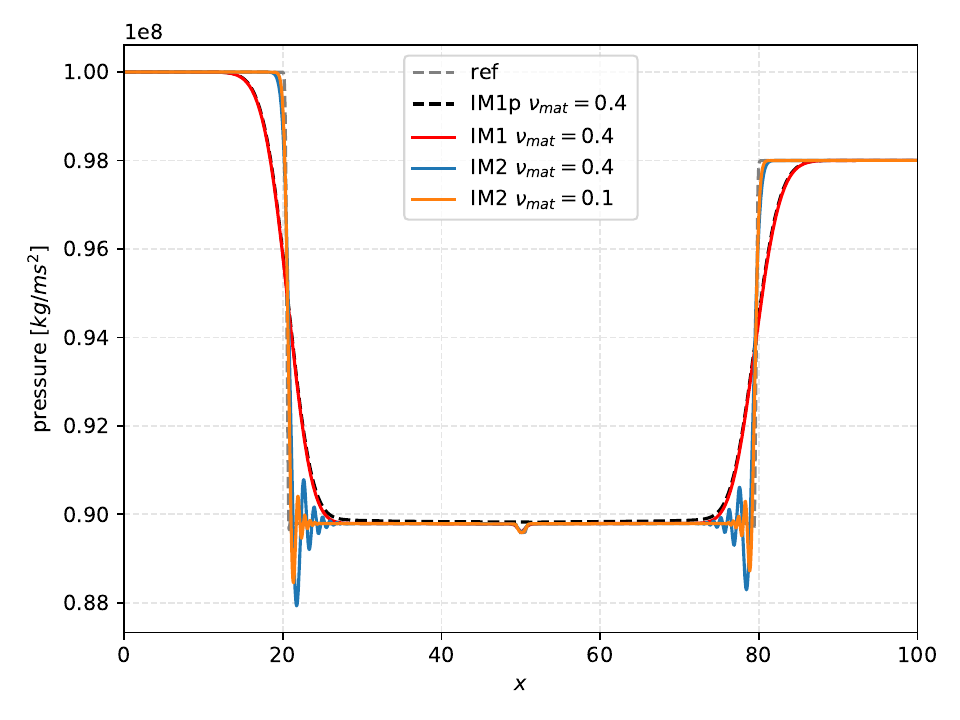}
            \includegraphics[scale=0.38]{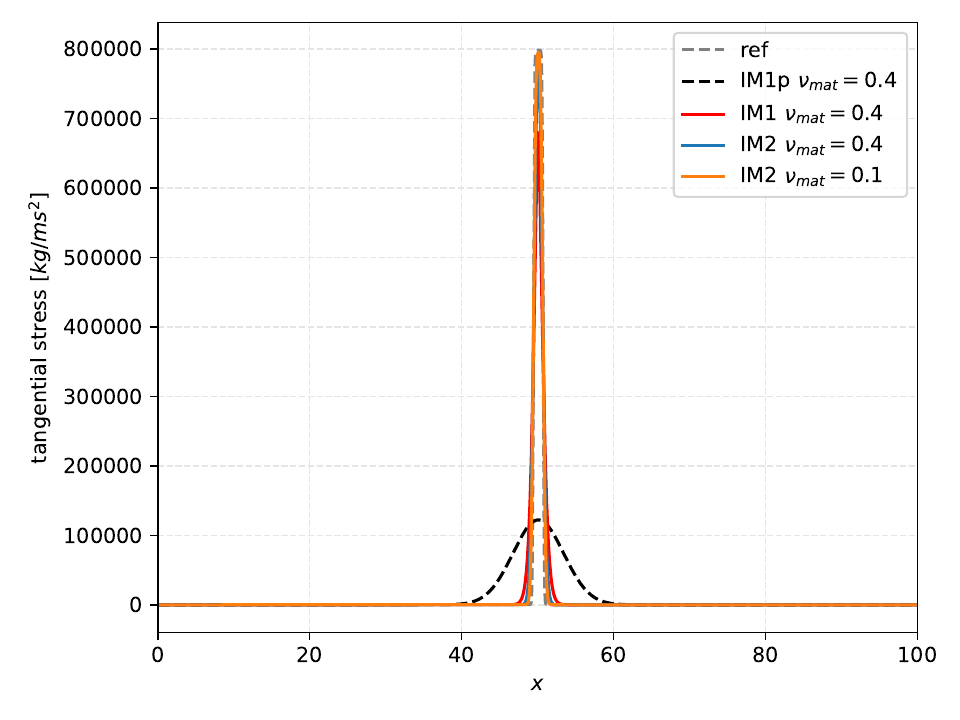}
        \end{center}
        \caption{Test 5: Density, velocity, pressure and tangential stress on computational domain $[0,100]$ with $\Delta x = 10^{-2}$ for the schemes IM1p, IM1 and IM2. }
        \label{fig:Test5full}
    \end{figure}

\begin{figure}[t!]
        \begin{center}
            \includegraphics[scale=0.397]{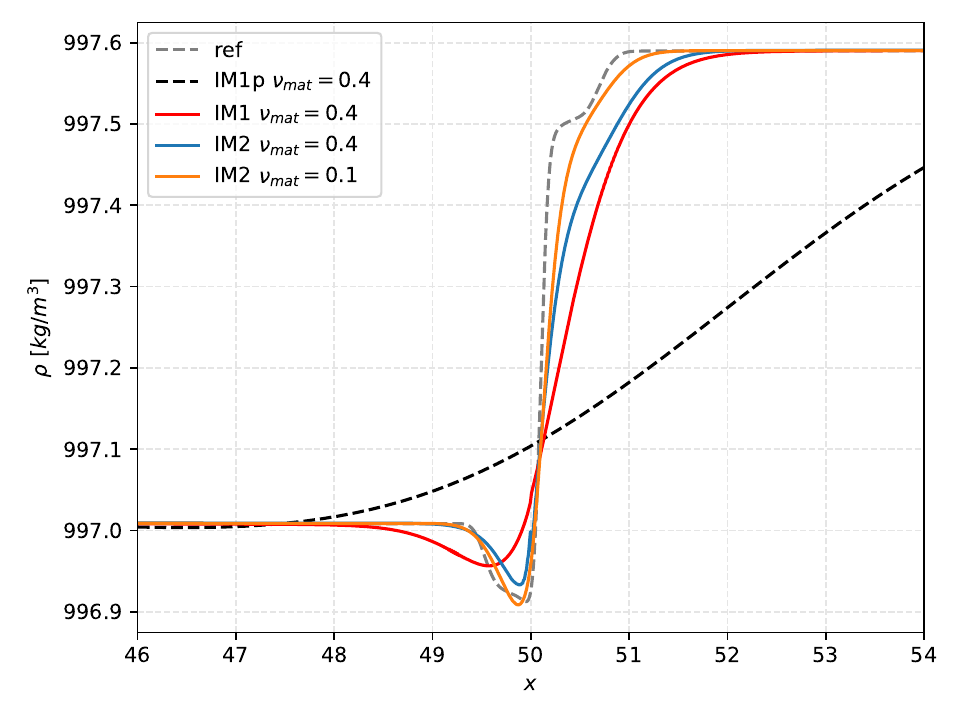}
            \includegraphics[scale=0.397]{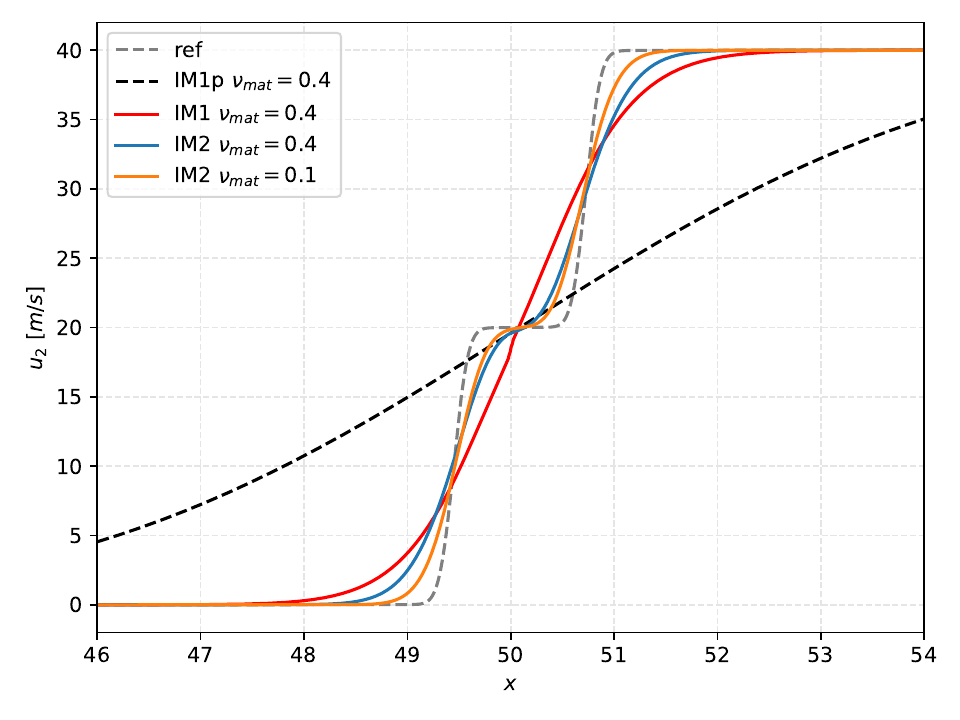}
            \includegraphics[scale=0.397]{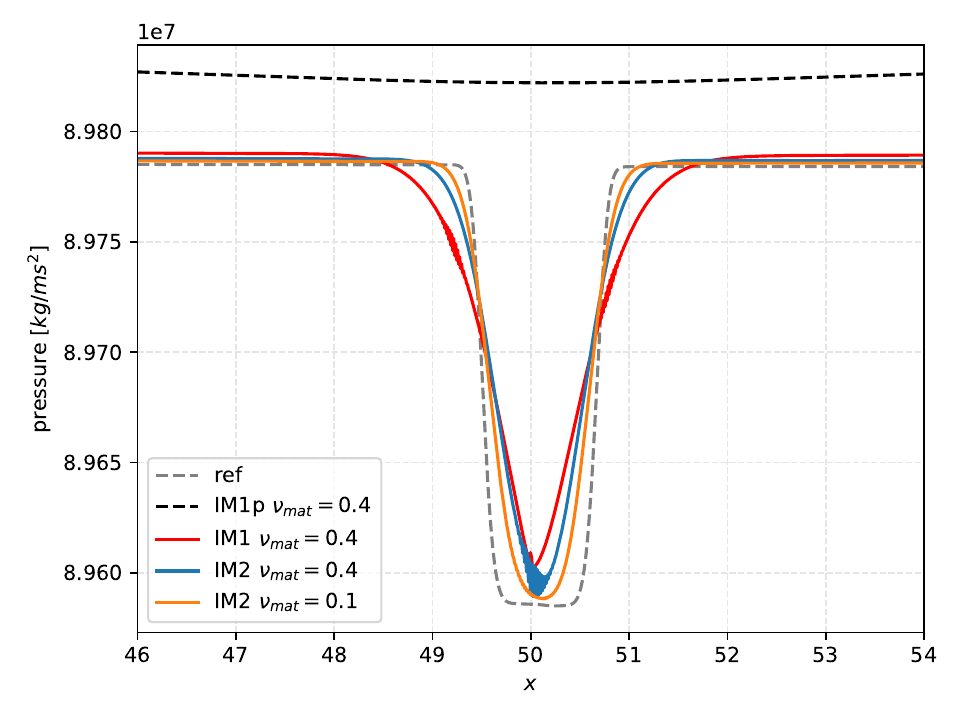}
            \includegraphics[scale=0.397]{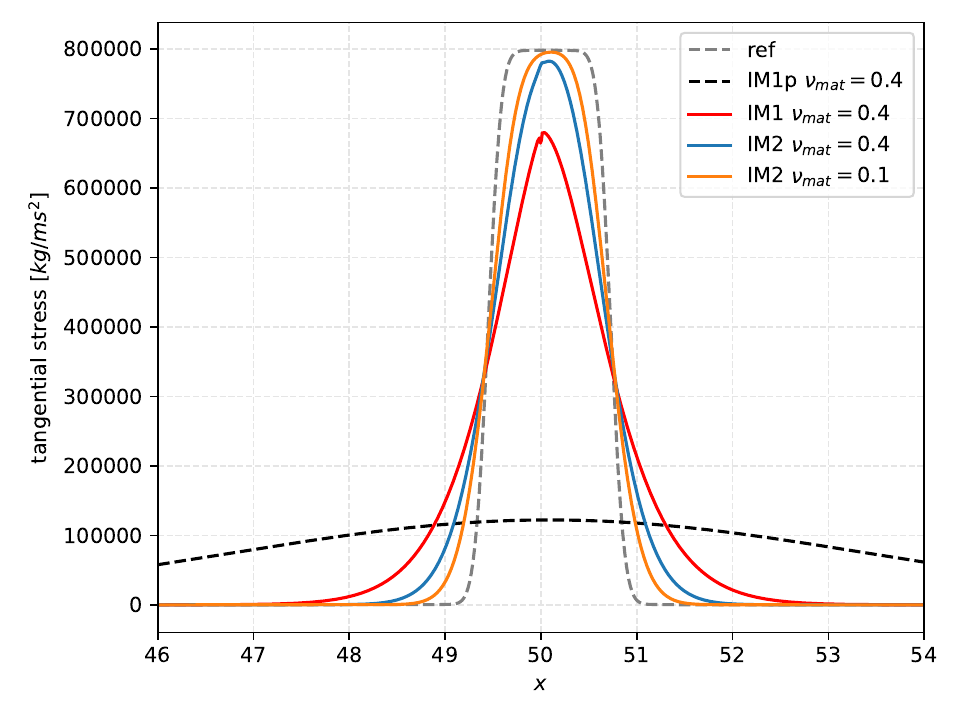}
        \end{center}
        \caption{Test 5: Zoom on the slow waves in the center of the computational domain $[0,100]$ with 10000 grid cells. }
        \label{fig:Test5}
    \end{figure}

    \textbf{Test 5} corresponds to an acoustic low Mach regime. 
    It simulates the deformation of a hyperelastic material (rubber) where the material parameter $p_\infty$ is several orders of magnitude larger than $\chi$, see Table \ref{tab:Parameters_mat_tests}. 
    The material and shear waves are almost stationary compared to the significantly faster longitudinal waves traveling towards the left and right boundaries of the domain. 
    The space discretization for all schemes are given by $N=10000$ grid cells resulting in $\Delta x = 10^{-2}$.  
    The time step for an explicit scheme would be constraint by the longitudinal waves with a CFL number given by 0.9 resulting in $\Delta t = 4.83\cdot10^{-6}$. 
    For the implicit schemes IM1 and IM2, larger time steps can be chosen given by $\Delta t = 1\cdot10^{-4}$ which corresponds to $\nu_{mat} = 0.4$ or respectively $\nu_{ac} = 18.75$ in the CFL condition. 
    The numerical results on the whole domain are given in Figure \ref{fig:Test5full}.
    To assess the influence of the time step on the oscillations on the fast longitudinal waves, we compare the results of scheme IM2 with $\nu_{mat}= 0.4$ with a smaller time step given by $\nu_{mat} = 0.1$. 
    We see from the results of Figure \ref{fig:Test5full}, that the oscillations consistently reduce with a smaller time step. 
    It is evident that the predictor scheme IM1p is too diffusive to resolve the slow waves.
    In Figure \ref{fig:Test5}, density, pressure and tangential stress computed with the schemes IM1 and IM2 are plotted with focus on the slow waves on the domain $[46,54]$.
    The material and shear waves are captured accurately with schemes IM1 and IM2 while the results with scheme IM1p are too diffusive to distinguish between the contact and shear waves in the density. 
    The reference solution in Figure \ref{fig:Test5full,fig:Test5} are computed with the IM2 scheme on a finer grid with $\Delta x = 2\cdot 10^{-3}$ and $\Delta t = 5\cdot 10^{-6}$. Using an explicit scheme, as done in Figure \ref{fig:Test41} for Test 4.1, even on fine grids still yields a too diffusive profile of the contact and shear waves and is therefore not feasible as reference solution.
    For a comparison of the results with the fully coupled scheme \eqref{eq:RelaxingScheme} given in \cite{AbbIolPup2017}, see Fig. 14 in \cite{AbbIolPup2017}.
    Note, that with our new decoupled approach, the computational time per cell is reduced by construction with respect to the previous work \cite{AbbIolPup2017} and thus allows a much better resolution in space without increasing the CPU times (2500 cells used in \cite{AbbIolPup2017} versus 10000 cells used here).
    This leads to a better capturing of the minimum in the pressure and maximum of the tangential stress in Figure \ref{fig:Test5}.
    
    \section{Conclusions and further developments}
    \label{sec:conclusions}
    
    We have proposed an implicit relaxed solver based on a Jin-Xin relaxation model for the numerical simulation of all-speed flows. 
    The scheme was constructed for general systems of hyperbolic conservation laws and was applied on the simulation of compressible materials using the Euler equations and a monolithic Eulerian model of non-linear elasticity. 
    The scheme has proved to be accurate on the approximation of material waves in different Mach regimes as well as on the computation of steady state solutions. 
    The presented scheme showed improvement over the implicit relaxing scheme introduced in \cite{AbbIolPup2017} due to the reduced number of variables that needed to be updated and the simple linear decoupled structure of the method. 
    However, in the context of Riemann problems, local oscillations on the acoustic waves in Figure \ref{fig:Test3} and longitudinal waves in Figure \ref{fig:Test41,fig:Test5full} appeared. 
    These phenomenon was also noted by the authors of \cite{PupSemVis2021} in the context of obtaining linear higher order space reconstruction methods coupled with SDIRK integrators in time for scalar equations. 
    The authors concluded, that a limiting in time is necessary to reduce those oscillations. 
    Although the oscillations we observed in our simulations were consistent, of local nature and appeared only on the negligible fast waves, it cannot be guaranteed that they might cause negative pressures or densities in different applications. 
    Therefore we plan to incorporate ideas on efficient time limiting proposed in \cite{PupSemVis2021} for systems in our linear implicit framework in future work.
    Another issue that presented itself during the simulation of long term monitoring of material waves as in Figure \ref{fig:Test42,fig:Test5full} is the incomplete knowledge of boundary conditions and is connected to relaxation model the scheme is based on. 
    Since the number of equations in the relaxation model are doubled with respect to the original equations, twice as many boundary conditions are necessary to describe the behavior of the medium at the boundaries.
    In the relaxation limit however, the number of characteristics reduces, but the nature of the problem changes from hyperbolic to parabolic as an elliptic operator appears for the state variables.
    To avoid the extension of the computational domain as done in Tests 4 and 5 for long term simulations, it would be preferable to have an appropriate description of the flow at the boundaries. This will be subject to future work.
	 
	\section*{Acknowledgements}
	A. T. has been partially supported by the Gutenberg Research College, JGU Mainz, and PROCOPE-MOBILITE-2021 granted by the \textit{Service pour la science et la technologie de l'Ambassade de France en Allemagne}.
    Further G. P. acknowledges the support of PRIN2017 and Sapienza, Progetto di Ateneo RM120172B41DBF3A.

	\bibliographystyle{plain}
	\bibliography{literature.bib}
\end{document}